\documentclass[preprints,article,accept,moreauthors,10pt,a4paper]{mdpi}

\firstpage{1}
\pubvolume{xx}
\issuenum{1}
\articlenumber{5}
\pubyear{2020}
\copyrightyear{2020}
\history{Received: date; Accepted: date; Published: date}
\usepackage{amssymb}

\usepackage{bm}
\usepackage{hyperref}
\usepackage[hyphenbreaks]{breakurl}
\usepackage{braket}
\usepackage[caption=false]{subfig}

\usepackage{environ}
\NewEnviron{myequation}{%
\begin{equation}
\scalebox{0.95}{$\BODY$}
\end{equation}
}
\NewEnviron{myequation2}{%
\begin{equation}
\scalebox{1}{$\BODY$}
\end{equation}
}
\NewEnviron{myequation3}{%
\begin{equation}
\scalebox{1}{$\BODY$}
\end{equation}
}
\NewEnviron{myequation4}{%
\begin{equation}
\scalebox{1}{$\BODY$}
\end{equation}
}

\newcommand{\dif}{\mathrm{d}}
\newcommand{\cc}{\bm{c}}
\newcommand{\vv}{\bm{v}}
\newcommand{\vva}{\bm{v}_{1}}
\newcommand{\vvb}{\bm{v}_{2}}
\newcommand{\vvab}{\bm{v}_{12}}
\newcommand{\s}{\widehat{\bm{\sigma}}}
\newcommand{\dt}{d}
\newcommand{\thr}{\text{th}}
\newcommand{\hcs}{\text{H}}
\newcommand{\refe}{\text{ref}}
\newcommand{\hh}{\Delta c}
\newcommand{\DKL}{\mathcal{D}_{\text{KL}}}

\def\bal#1\eal{\begin{align}#1\end{align}}
\newcommand\BEQ{\begin{equation}}
\newcommand\EEQ{\end{equation}}
\newcommand\beQa{\begin{eqnarray}}
\newcommand\eeQa{\end{eqnarray}}
\newcommand{\nn}{\nonumber\\}
\newcommand{\LL}[1]{\mathcal{L}_{3}\left\{#1\right\}}
\newcommand{\Ll}[1]{\mathcal{L}_{2}\left\{#1\right\}}
\newcommand{\llangle}{\langle\!\langle}
\newcommand{\rrangle}{\rangle\!\rangle}
\Title{Kullback--Leibler Divergence of a Freely Cooling Granular Gas}
\Author{%
{Alberto} %Please carefully check the accuracy of names and affiliations.
 Meg\'ias $^{1}$\orcidB{} and Andr\'es Santos $^{2,}$*\orcidA{}}

\AuthorNames{Alberto Meg\'ias and Andr\'es Santos}

\address{%
$^{1}$ \quad Departamento de F\'isica, Universidad de
Extremadura, E-06006 Badajoz, Spain; albertom@unex.es\\
$^{2}$ \quad Departamento de F\'isica and Instituto de Computaci\'on Cient\'ifica Avanzada (ICCAEx), \mbox{Universidad de
Extremadura}, E-06006 Badajoz, Spain}

\corres{Correspondence: andres@unex.es; Tel.: +34-924-289-651}

\abstract{Finding the proper entropy-like Lyapunov functional associated with the inelastic Boltzmann equation for an isolated freely cooling granular gas is a still unsolved challenge. The~original $H$-theorem hypotheses do not fit here and the $H$-functional presents some additional measure problems that are solved by the Kullback--Leibler divergence (KLD) of a reference velocity distribution function from the actual distribution. The right choice of the reference distribution in the KLD is crucial for the latter to qualify or not as a Lyapunov functional,  the asymptotic ``homogeneous cooling state'' (HCS) distribution  being a potential candidate. Due to the lack of a formal proof far from the quasielastic limit, the aim of this work is to support this conjecture aided by molecular dynamics simulations of inelastic hard disks and spheres in a wide range of values for the coefficient of restitution ($\alpha$) and for different initial conditions. Our results reject the Maxwellian distribution as a possible reference, whereas they reinforce the HCS one. Moreover, the KLD is used to measure the amount of information lost on using the former rather than the latter, revealing a non-monotonic dependence with $\alpha$. }

\keyword{Kullback--Leibler divergence; granular gases; kinetic theory; molecular dynamics}

\begin{document}
%%%%%%%%%%%%%%%%%%%%%%%%%%%%%%%%%%%%%%%%%%

%%%%%%%%%%%%%%%%%%%%%%%%%%%%%%%%%%%%%%%%%%

%\section{How to Use this Template}
%The template details the sections that can be used in a manuscript. Note that the order and names of article sections may differ from the requirements of the journal (e.g., the positioning of the Materials and Methods section). Please check the instructions for authors page of the journal to verify the correct order and names. For any questions, please contact the editorial office of the journal or support@mdpi.com. For LaTeX related questions please contact latex@mdpi.com.
%The order of the section titles is: Introduction, Materials and Methods, Results, Discussion, Conclusions for these journals: aerospace,algorithms,antibodies,antioxidants,atmosphere,axioms,biomedicines,carbon,crystals,designs,diagnostics,environments,fermentation,fluids,forests,fractalfract,informatics,information,inventions,jfmk,jrfm,lubricants,neonatalscreening,neuroglia,particles,pharmaceutics,polymers,processes,technologies,viruses,vision

\section{Introduction}

Thermodynamics and information theory are clearly connected via the entropy concept. This~idea allows physicists to understand plenty of details and consequences in the evolution and intrinsic behavior of physical systems.
 {However}%Authors: Merging of both paragraphs
, finding the entropy-like Lyapunov functional for a given problem is not an easy task. Thankfully, information theory provides tools that one can use in physics problems, usually proving a rewarding~feedback.
 {Along this paper}%Authors: Parentheses removed and sentence moved
, and~as usually done in the context of information theory~\cite{S48,G11b} and nonequilibrium statistical mechanics~\cite{BS92,K14}, we borrow from equilibrium statistical mechanics and thermodynamics the use of the term ``entropy'' in a broader sense. The~same applies to the term ``temperature,'' introduced in Equation \eqref{eq:T_g} below.

In this work, we address the quest of  finding  the  Lyapunov  functional  of an isolated freely cooling monodisperse granular gas, modeled by identical inelastic and smooth hard disks ($d=2$) or hard spheres ($d=3$) with constant coefficient of restitution ($\alpha$). The~interest of this study does not only reside in the mathematical challenge, but~also in the physical consequences for granular matter. Typically, for~a classical gas, Boltzmann's $H$-theorem provides the desired entropy-like Lyapunov functional~\cite{CC70,GS03}. Nevertheless, inelasticity plays a fundamental role in the dynamics, and~the  hypotheses of the latter theorem are not applicable. Previous works have proposed the Kullback--Leibler  divergence \mbox{(KLD) \cite{KL51,K78}} as the proper alternative to the $H$-functional~\cite{SK12,BPV13,GMMMRT15,PP17}. One of the aims of this paper is to explore with molecular dynamics (MD) simulations~\cite{BSL11} the validity  of the KLD as a Lyapunov  functional  in the whole range of definition of $\alpha$ and for both disks and~spheres.

The freely cooling one-particle velocity distribution function (VDF) of our granular-gas model is expected to asymptotically reach a scaled form,  the~so-called ``homogenous cooling state'' (HCS), $f_{\text{HCS}}$. Although~its explicit form is unknown, there is a vast amount of literature about it~\cite{G19,BP04,BP00,vNE98,MS00,SM09,BRC96,AP06,AP07} and recent experiments have demonstrated some of their properties~\cite{YSS20}. While computational and experimental evidence supporting the HCS are overwhelming, a~rigorous mathematical proof on its existence and long-time approach has only been achieved for inelastic Maxwell models described by the Boltzmann equation~\cite{BCT03,BCT06,BC07,CT07,CCC09}.

The HCS VDF $f_{\text{HCS}}$ is usually expressed as an infinite expansion around the Maxwellian VDF in terms of Sonine polynomials~\cite{CC70,G19,BP04}, even though the expansion may break down for large  {inelasticities}%Authors: Old Ref. [30] merged into Ref.[29]
~\cite{BP06,NBSG07}.
Here, in~order to provide a detailed description of the problem for both the stationary and transient regimes, we revisit some well-known results and also provide new simulation data and theoretical expressions obtained from a truncation in the Sonine expansion up to the sixth cumulant.
In particular, our MD simulation results for the HCS fourth and sixth cumulants are compared with previous ``direct simulation Monte Carlo''  (DSMC) results~\cite{MS00,SM09} and a good agreement is~found.

The paper is structured as follows. In~Section~\ref{sec2}, the~Sonine expansion formalism is presented and  simulation and theoretical results for the fourth and sixth cumulants are provided.
The measure problem introduced by the original $H$-functional is established in Section~\ref{sec3} and the KLD for  two different reference VDFs  is studied and compared with MD simulation outcomes.  Finally, in~Section~\ref{sec5}, some concluding remarks of this work are presented and~discussed.

%%%%%%%%%%%%%%%%%%%%%%%%%%%%%%%%%%%%%%%%%%
\section{Free Cooling Evolution of Velocity~Cumulants}
\label{sec2}
\vspace{-6pt}
\subsection{Boltzmann Equation and~HCS}
\label{sec2.1}

Consider a model of a monodisperse granular gas consisting of an isolated  collection of inelastic hard $d$-spheres of mass $m$, diameter $\sigma$, and~a constant  coefficient of normal restitution $\alpha<1$.
Under the molecular chaos ansatz (\emph{Stosszahlansatz}), the~free cooling of a homogeneous and isotropic gas can be described by the Boltzmann equation~\cite{G19}
\begin{myequation}
\label{EB}
\partial_t f(\vva;t)=n\sigma^{d-1}I[\vva|f,f]\equiv n\sigma^{d-1}\int \dif\vvb\int_+ \dif\s\, (\vvab\cdot\s)\left[\alpha^{-2}f(\vva'';t)f(\vvb'';t)-f(\vva;t)f(\vvb;t)\right],
\end{myequation}
where $n$ is the number density,  $\vvab=\vva-\vvb$ is the relative velocity of the two colliding particles, $\s$ is a unit vector along the line of centers from particle $1$ to particle $2$,
the subscript $+$ in the integral over $\s$ means the constraint  $\vvab\cdot\s>0$, and~\begin{equation}
\label{coll_rule}
\vva''=\vva-\frac{1+\alpha}{2\alpha}(\vvab\cdot\s)\s,\quad \vvb''=\vvb+\frac{1+\alpha}{2\alpha}(\vvab\cdot\s)\s
\end{equation}
are precollisional velocities. Note that we have defined the VDF with the normalization condition $\int\dif\vv \,f(\vv;t)=1$.

An important quantity is the \emph{granular} temperature defined as
\begin{equation}\label{eq:T_g}
    T(t) = \frac{m}{d}\braket{v^2},\quad \braket{X(\vv)}\equiv \int\dif\vv \, X(\vv)f(\vv;t).
\end{equation}
Taking moments in Equation \eqref{EB}, one finds the cooling equation
\begin{equation}
\label{dotT}
\partial_t{T}(t)=-\zeta(t)T(t),
\end{equation}
where the cooling rate is given by
\begin{subequations}
\begin{equation}
\label{zeta}
\zeta(t)=-\frac{mn\sigma^{d-1}}{T(t)d}\int\dif\vv\,v^2 I[\vv|f,f]=(1-\alpha^2)\frac{mn\sigma^{d-1}}{T(t)}\frac{\pi^{(d-1)/2}}{4d\Gamma(\frac{d+3}{2})}\llangle{v_{12}^3}\rrangle,
\end{equation}
\begin{equation}
\llangle{X(\vva,\vvb)}\rrangle\equiv \int\dif\vva\int\dif\vvb \,X(\vva,\vvb)f(\vva;t)f(\vvb;t).
\end{equation}
\end{subequations}

Let us introduce the \emph{thermal} velocity $v_{\thr}(t)\equiv \sqrt{2T(t)/m}$, which allows us to define the  {rescaled} %Authors: Italics removed
 VDF $\phi(\cc;s)$ as
\begin{equation}
\label{eq:phi}
    f(\vv;t)=v_{\thr}^{-d}(t)\phi(\cc;s),\quad \cc\equiv\frac{\vv}{v_\thr(t)},
\end{equation}
where the variable $s$ in $\phi(\cc;s)$ is a scaled time defined by
\begin{equation}
\label{s(t)}
    s(t) = \frac{1}{2}\int_0^t \dif t^\prime\, \nu(t^\prime),\quad \nu(t)\equiv {\kappa}{n\sigma^{d-1}v_{\thr}(t)},\quad \kappa\equiv \frac{\sqrt{2}\pi^{(d-1)/2}}{\Gamma(\frac{d}{2})}.
\end{equation}
Here, $\nu$ is the (nominal) collision frequency, so that $s(t)$ represents the (nominal) accumulated average number of collisions per particle up to time $t$.
In terms of these dimensionless quantities, the~Boltzmann Equation \eqref{EB} can be rewritten as
\begin{equation}
\label{BE_phi}
\frac{\kappa}{2}\partial_s\phi(\cc;s)+\frac{\mu_2(s)}{d}\frac{\partial}{\partial \cc}\cdot\left[\cc \phi(\cc;s)\right]=I[\cc|\phi,\phi],\quad  \mu_k(s)\equiv-
\int\dif\cc\,c^k I[\cc|\phi,\phi],
\end{equation}
where we have taken into account that $\zeta(t)/n\sigma^{d-1}v_\thr(t)={2}\mu_2(s)/{d}$. The~associated hierarchy of moment equations is
\begin{equation}
\label{moments}
\frac{\kappa}{2}\partial_s\braket{c^k}=F_k(s)\equiv \frac{k\mu_2(s)}{d}\braket{c^k}-\mu_k(s).
\end{equation}
Note that $F_0=F_2=0$, since $\mu_0=0$ and $\braket{c^2}=\frac{d}{2}$.

In the long-time limit, the~free cooling is expected to reach an asymptotic regime (the HCS) in which the scaled VCF is  {stationary} %Authors: Italics removed
, i.e.,~$\phi(\cc;s)\to \phi_\hcs(\cc)$, where $\phi_\hcs(\cc)$ satisfies the integrodifferential equation
\begin{equation}
\label{BE_HCS}
\frac{\mu_2^\hcs}{d}\frac{\partial}{\partial \cc}\cdot\left[\cc \phi_\hcs(\cc)\right]=I[\cc|\phi_\hcs,\phi_\hcs].
\end{equation}
 {Henceforth,} %Authors: Parentheses removed and sentence moved
  a~subscript or superscript $\hcs$ on a quantity means that the quantity is evaluated in the HCS.
Within that regime, Equation \eqref{zeta} shows that $\zeta_\hcs(t)/\sqrt{T_\hcs(t)}=\text{const}$, so that the solution to Equation~\eqref{dotT} gives rise to the well-known cooling Haff's law~\cite{G19,BP04,BE98}
\begin{equation}\label{eq:Haff_t}
    T_\hcs(t) = \frac{T_\hcs(t_0)}{\left[1+\frac{1}{2}\zeta_\hcs(t_0)(t-t_0)\right]^2},
\end{equation}
$t_0$ being an arbitrary time belonging to the HCS regime. Also in the HCS regime, $\mu_2(s)\to\mu_2^\hcs=\text{const}$ and thus Equation \eqref{dotT} becomes $\partial_s T_\hcs(s)=-(4/\kappa d)\mu_2^\hcs T_\hcs(s)$, whose solution is
\begin{equation}\label{eq:Haff}
    T_\hcs(s)= T_\hcs(s_0) e^{-4\mu_2^\hcs(s-s_0)/\kappa d}.
\end{equation}
Therefore, in~the HCS, the temperature decays exponentially with the average number of collisions per~particle.

\subsection{Sonine Expansion~Formalism}
\label{sec2.2}

The Maxwell--Boltzmann VDF $\phi_{\text{M}}(\cc)=\pi^{-d/2}e^{-c^2}$ is not a solution of the HCS Boltzmann Equation \eqref{BE_HCS}. While its analytic form  has not been found, the~HCS solution is known to be rather close to $\phi_{\text{M}}$ in the domain of thermal velocities (\emph{c}$\sim$1) \cite{BRC96}. Thus, it is convenient to represent the time-dependent VDF in terms of a Sonine polynomial expansion,
\begin{equation}
\label{eq:fv}
    \phi(\cc;s)= \phi_{\text{M}}(\cc)\left[1+\sum_{k=2}^\infty a_k(s) S_k(c^2) \right],
\end{equation}
where
\begin{equation}
S_k(x)=L_k^{(\frac{d}{2}-1)}(x)=\sum_{j=0}^k \frac{(-1)^j\Gamma\left(\frac{d}{2}+k\right)}{\Gamma\left(\frac{d}{2}+j\right)(k-j)!j!}x^j
\end{equation}
are Sonine (or generalized Laguerre) polynomials, which satisfy the orthogonalization condition
\begin{equation}
\label{ortho}
     \braket{S_k | S_{k^\prime}}\equiv \int \dif\cc \medspace\phi_{\text{M}}(\cc) S_k(c^2)S_{k^\prime}(c^2)=\mathcal{N}_{k}\delta_{k,k^\prime},\quad \mathcal{N}_{k}\equiv \frac{\Gamma\left(\frac{\dt}{2}+k\right)}{\Gamma\left(\frac{d}{2}\right)k!}.
\end{equation}
In Equation \eqref{eq:fv}, the~Sonine coefficient $a_k(s)$ is the $2k$-th cumulant of the VDF at time $s$. According to Equation \eqref{ortho},
\begin{equation}
    a_k(s)=\frac{\braket{S_k(c^2)}}{\mathcal{N}_k}.
\end{equation}
In particular, $a_0(s)=1$, $a_1(s)=0$, and~\begin{align}
\label{eq:a2_a3}
    a_2(s) = \frac{4}{\dt(\dt+2)}\langle c^4\rangle -1, \quad a_3(s)= 1+3a_2-\frac{8}{\dt(\dt+2)(\dt+4)}\langle c^6\rangle.
\end{align}

\subsection{Truncated Sonine~Approximation}
\label{sec2.3}

Thus far, all the results presented in Sections~\ref{sec2.1} and \ref{sec2.2} are formally exact within the framework of the homogeneous Boltzmann Equation~\eqref{EB}. However, in~order to obtain explicit results, we need to resort to~approximations.

As usual~\cite{GS95,vNE98,MS00,BP00,BP04,BP06,SM09}, we will start by neglecting the coefficients $a_k$ with $k\geq 4$ in Equation~\eqref{eq:fv}, as~well as the nonlinear terms  $a_2^2$, $a_2a_3$, and~$a_3^2$ in the bilinear collision operator $I[\cc|\phi,\phi]$.
Given~a functional $X[\phi]$ of the scaled VDF $\phi(\cc)$, we will use the notation
$\LL{X}$ to denote the result of that truncation and linearization procedure. Furthermore, if~$a_3$ is also
neglected, the~corresponding approximation will be denoted by
$\Ll{X}$. In~particular, in~the case of the collisional moments $\mu_2$, $\mu_4$, and~$\mu_6$, one has
\begin{equation}
\LL{\mu_2}=
A_0+A_2a_2+A_3a_3,\quad \LL{\mu_4}=
B_0+B_2a_2+B_3a_3,\quad \LL{\mu_6}=
C_0+C_2a_2+C_3a_3,
\label{15}
\end{equation}
where the expressions for the coefficients $A_i$, $B_i$, and~$C_i$ as
functions of $\alpha$ and $d$ can be found in  {\mbox{Ref.~\cite{BP06}}} %Authors: Old Ref. [30] merged into Ref.[29]
 and in  Appendix A of Ref.\ \cite{SM09}.
Obviously, $\Ll{\mu_2}$,  $\Ll{\mu_4}$, and~$\Ll{\mu_6}$ are obtained by formally setting $A_3\to 0$, $B_3\to 0$, and~$C_3\to 0$, respectively.

Let us first use the simple approximation $\mathcal{L}_2$ to estimate $a_2^\hcs$. From~Equation \eqref{moments}, we have that $F_4^\hcs=0$. Thus, the obvious approximation~\cite{vNE98} consists of
\begin{equation}
\label{a2a}
\Ll{F_4^\hcs}=0\Rightarrow a_2^{\hcs,a}=\frac{(d+2)A_0-B_0}{B_2-(d+2)(A_2+A_0)}=
\frac{16(1-\alpha)(1-2\alpha^2)}{9+24d-(41-8
d)\alpha+30(1-\alpha)\alpha^2},
\end{equation}
where, in the last steps, use has been made of the explicit
expressions of $A_0$, $A_2$, $B_0$, and~$B_2$.
However, this is not by any means the only possibility of estimating $a_2^\hcs$ \cite{MS00,CDPT03,SM09}. In~particular, one can start from the logarithmic time derivative of the fourth moment and then take
\begin{equation}
\label{a2b}
\Ll{\frac{F_4^\hcs}{\braket{c^4}_\hcs}}=0\Rightarrow a_2^{\hcs,b}=\frac{(d+2)A_0-B_0}{B_2-B_0-(d+2)A_2}=
\frac{16(1-\alpha)(1-2\alpha^2)}{25+24d- (57-
8d)\alpha-2(1-\alpha)\alpha^2}.
\end{equation}
Note that
\begin{equation}
\frac{a_2^{\hcs,a}}{a_2^{\hcs,b}}=1+a_2^{\hcs,a}=\frac{1}{1-a_2^{\hcs,b}}.
\end{equation}
Both approximations ($a_2^{\hcs,a}$ and $a_2^{\hcs,b}$) are practically indistinguishable in the region $0.6\lesssim \alpha<1$, but~$a_2^{\hcs,b}$ is much more accurate than $a_2^{\hcs,a}$ for higher inelasticity~\cite{MS00,SM09}.

Next, to~estimate $a_3^\hcs$, we start from the exact condition $F_6^\hcs=0$ and carry out either the linearization
\begin{myequation2}
\LL{F_6^\hcs}=0\Rightarrow a_3^{\hcs,a}=G_a(a_2^\hcs)\equiv
\frac{C_0-\frac{3}{4}(d+2)(d+4)A_0+\left[C_2-\frac{3}{4}(d+2)(d+4)(3A_0+A_2)\right]a_2^\hcs}{\frac{3}{4}(d+2)(d+4)(A_3-A_0)-C_3}
\label{n10}
\end{myequation2}
or, alternatively,
\begin{myequation3}
\LL{\frac{F_6^\hcs}{\braket{c^6}_\hcs}}=0\Rightarrow a_3^{\hcs,b}=G_b(a_2^\hcs)\equiv
\frac{C_0-\frac{3}{4}(d+2)(d+4)A_0+\left[C_2-3C_0-\frac{3}{4}(d+2)
(d+4)A_2\right]a_2^\hcs}{\frac{3}{4}(d+2)(d+4)A_3-C_3-C_0}.
\label{n11}
\end{myequation3}
In Equations \eqref{n10} and \eqref{n11}, $a_3^\hcs$ is expressed in terms of $a_2^\hcs$. Using Equations \eqref{a2a} and \eqref{a2b}, four~possibilities in principle arise, namely
\begin{equation}
\label{a3ab}
a_3^{\hcs,aa}=G_a(a_2^{\hcs,a}),\quad a_3^{\hcs,ab}=G_a(a_2^{\hcs,b}),\quad a_3^{\hcs,ba}=G_b(a_2^{\hcs,a}),\quad a_3^{\hcs,bb}=G_b(a_2^{\hcs,b}).
\end{equation}
Comparison with  DSMC results shows that the best general estimates are provided by $a_3^{\hcs,aa}$ and $a_3^{\hcs,ab}$.
In what follows, we choose $a_2^{\hcs,b}$ for the fourth cumulant and, for~the sake of consistency with that choice, we adopt $a_3^{\hcs,ab}$ for the sixth cumulant.
To simplify the notation, we make $a_2^{\hcs,b}\to a_2^\hcs$ and $a_3^{\hcs,ab}\to a_3^\hcs$.

Once the (approximate) HCS values $a_2^\hcs$ and $a_3^\hcs$ have been obtained, we turn our attention to the evolution equations of $a_2(s)$ and $a_3(s)$. Approximating Equation \eqref{moments} with $k=4$ as $\frac{\kappa}{2}\partial_s \ln\braket{c^4}=\Ll{F_4(s)/\braket{c^4}}$, one obtains
\begin{equation}
\label{dsa2}
\partial_s a_2(s)=-K_2\left[1+a_2(s)\right]\left[a_2(s)-a_2^{\hcs}\right],\quad K_2\equiv \frac{8}{d(d+2)\kappa}\left[B_2-B_0-(d+2)A_2\right].
\end{equation}
Its solution is
\begin{equation}
\label{Sol_a2}
a_2(s)=a_2^{\hcs}+\frac{1+a_2^{\hcs}}{X_0 e^{\gamma s}-1},\quad X_0\equiv \frac{1+a_2(0)}{a_2(0)-a_2^{\hcs}},\quad \gamma\equiv \left(1+a_2^{\hcs}\right)K_2.
\end{equation}
Analogously, if~Equation \eqref{moments} with $k=6$ is approximated as $\frac{\kappa}{2}\partial_s \braket{c^6}=\LL{F_6(s)}$, the~resulting evolution equation for $a_3$ is
\begin{equation}
\label{dsa3}
\partial_s a_3(s)=3\partial_s a_2(s)-K_2'\left[a_2(s)-a_2^{\hcs}\right]-K_3\left[a_3(s)-a_3^{\hcs}\right],
\end{equation}
where
\begin{subequations}
\begin{equation}
K_2'\equiv \frac{16}{d(d+2)(d+4)\kappa}\left[\frac{3}{4}(d+2)(d+4)(A_2+3A_0)-C_2\right],
\end{equation}
\begin{equation}
K_3\equiv \frac{16}{d(d+2)(d+4)\kappa}\left[\frac{3}{4}(d+2)(d+4)(A_3-A_0)-C_3\right].
\end{equation}
\end{subequations}
Taking into account Equation \eqref{Sol_a2}, the~solution to Equation \eqref{dsa3} is
\begin{subequations}
\label{Sol_a3}
\begin{equation}
a_3(s)=a_3^\hcs+Y_0e^{-K_3 s}+\left(1+a_2^\hcs\right)\left[\frac{3}{X_0e^{\gamma s}-1}+\left(\frac{K_2'}{K_3}+3\right)\,_2F_1\left(1,\frac{K_3}{\gamma};\frac{K_3}{\gamma}+1;X_0 e^{\gamma s}\right)\right],
\end{equation}
\begin{equation}
Y_0\equiv a_3(0)-a_3^\hcs-\left(1+a_2^\hcs\right)\left[\frac{3}{X_0-1}+\left(\frac{K_2'}{K_3}+3\right)\,_2F_1\left(1,\frac{K_3}{\gamma};\frac{K_3}{\gamma}+1;X_0 \right)\right],
\end{equation}
\end{subequations}
where $_2F_1(a,b;c;z)$ is the hypergeometric function~\cite{AS72}.

As far as we know, Equations \eqref{Sol_a2} and \eqref{Sol_a3} had not been obtained~before.

\subsection{Comparison with MD~Simulations}
The approximate theoretical predictions for $a_2^\hcs$ and $a_3^\hcs$ were tested against results obtained from the DSMC simulation method in, for~instance, Refs.\ \cite{BRC96,MS00,SM09}. However, since the DSMC method is a stochastic scheme to numerically solve the Boltzmann equation~\cite{B94}, it does not prejudice by construction the hypotheses upon which the Boltzmann equation is derived, in~particular the molecular chaos ansatz. Therefore, it seems important to validate the Sonine approximations for $a_2^\hcs$ and $a_3^\hcs$ by event-driven MD simulations  as well. In addition, the~theory allows us to solve the initial-value problem  and predict the evolution of the fourth and sixth cumulants, as~shown by Equations \eqref{Sol_a2} and \eqref{Sol_a3}, and~an assessment of those solutions is in~order.

In our MD simulations,  we studied systems with densities $n\sigma^d\approx 5\times 10^{-4}$ and $2\times 10^{-4}$ for disks and spheres, respectively.
It is known that the HCS exhibits a shearing/clustering instability for sufficiently large systems~\cite{BDKS98,G19}. To~prevent this, the~side length of the simulation box was chosen as $L/\sigma\approx 5\times 10^3$ for disks and $L/\sigma\approx 4\times 10^2$ for spheres (see Appendix \ref{app:A} for technical details). These~values are about $2$ and $30$ times smaller, respectively, than~the critical values beyond which the HCS becomes unstable  in the less favorable case considered ($\alpha=0.1$). Moreover, we have expressly verified that the systems remain stably homogeneous even for long~times.

\begin{figure}[H]
\centering
\includegraphics[width=.45\columnwidth]{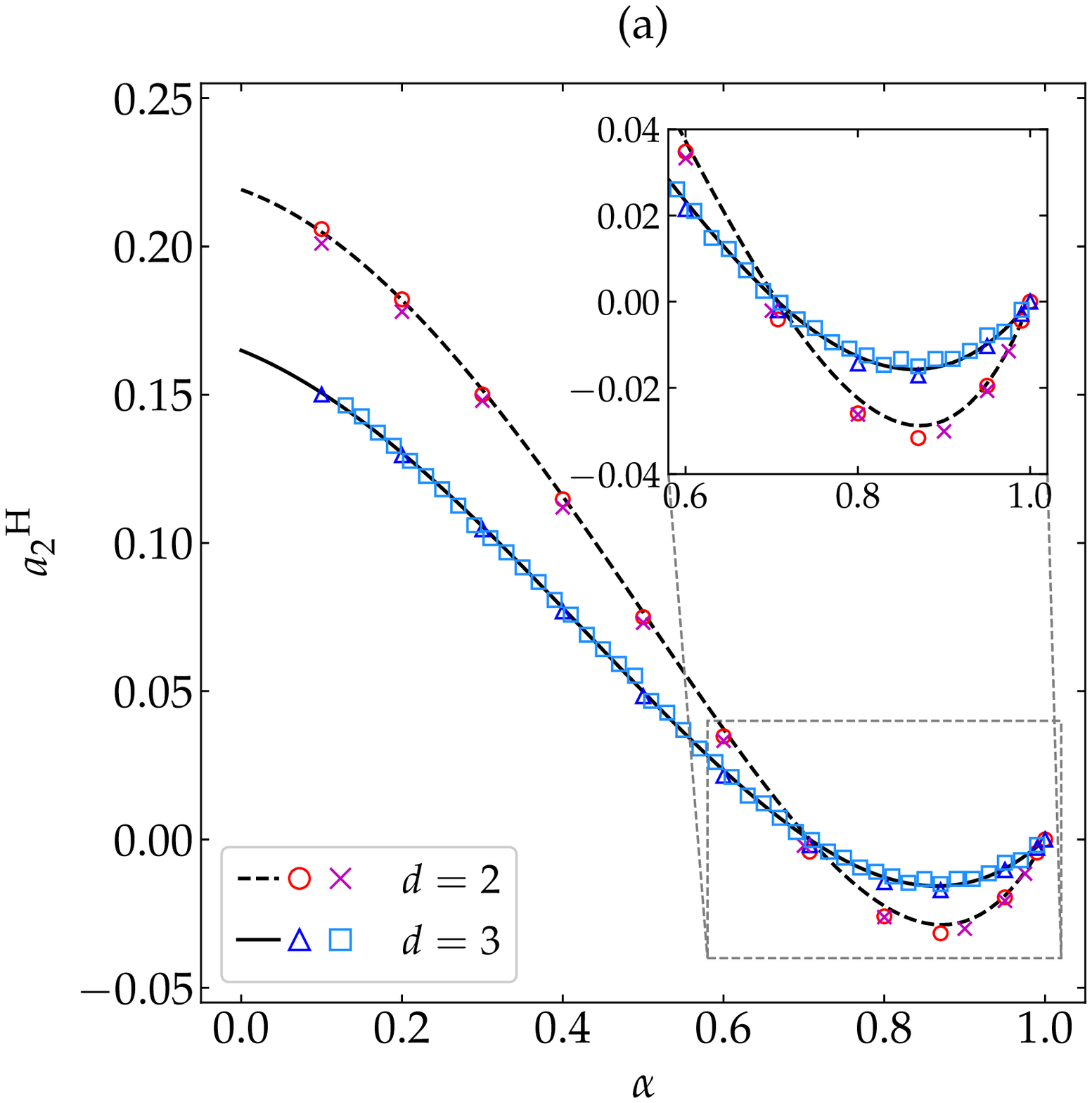}\hspace{1cm}\includegraphics[width=.45\columnwidth]{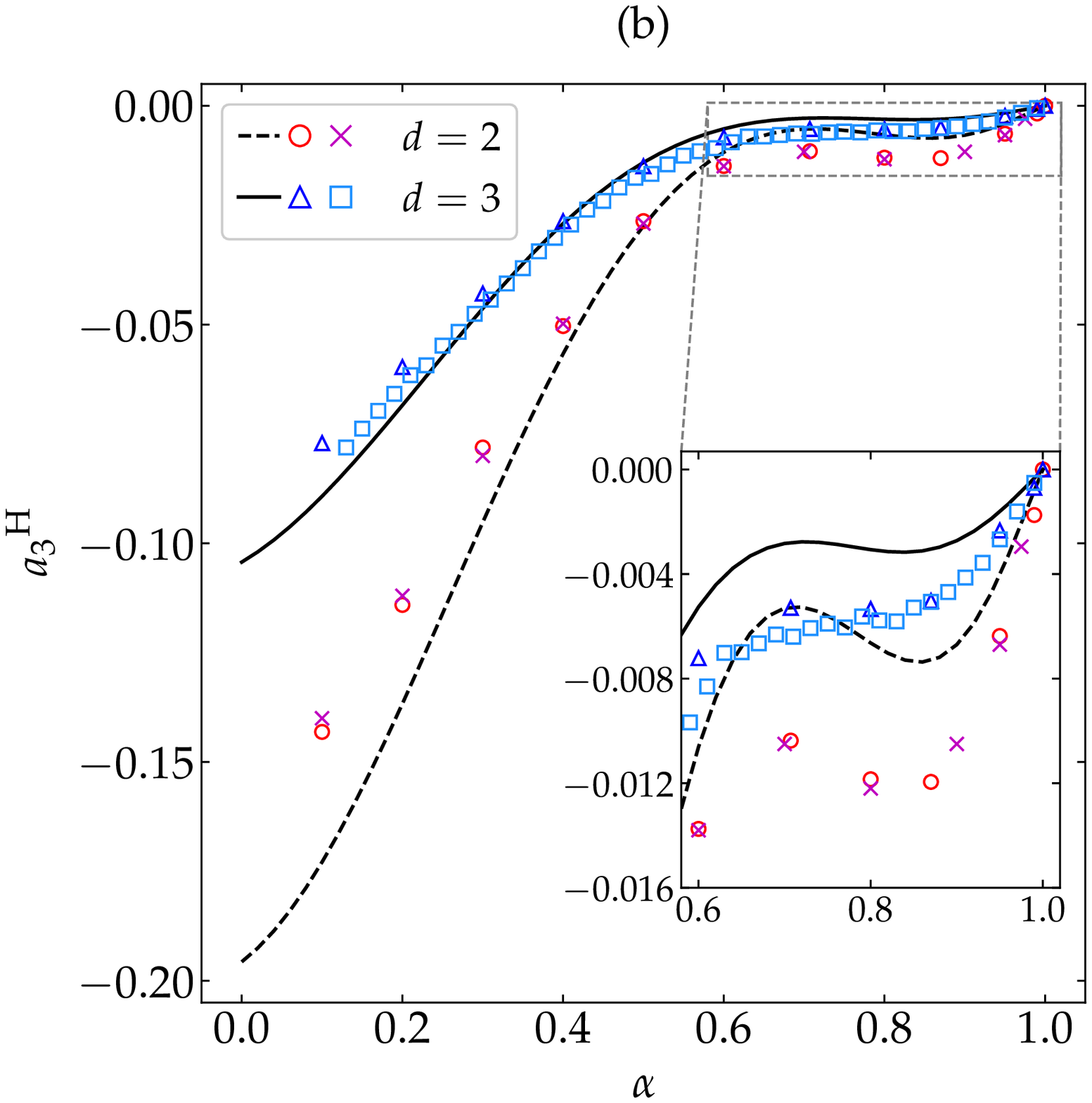}
\caption{Plot of (\textbf{a}) the HCS fourth cumulant $a_2^\hcs$ and (\textbf{b}) the HCS sixth cumulant $a_3^\hcs$ versus the coefficient of  restitution $\alpha$. Symbols represent simulation results: MD (this work) for disks ($\circ$) and spheres ($\vartriangle$), and~DSMC~\cite{MS00,BP06,SM09} for disks ($\times$) and spheres ($\square$). The~lines are the theoretical predictions $a_2^{\hcs,b}$ (see Equation \eqref{a2b}) and $a_3^{\hcs,ab}$ (see Equations \eqref{n10} and \eqref{a3ab}). The~insets magnify the
region \mbox{$0.6\leq\alpha\leq 1$}. The~error bars in the simulation data are smaller than the size of the symbols.\label{fig:a2a3st}}
\end{figure}

Figures~\ref{fig:a2a3st}a,b show the $\alpha$-dependence of $a_2^\hcs$ and $a_3^\hcs$, respectively, for~both hard disks ($d=2$) and spheres ($d=3$). An~excellent agreement between the MD
and DSMC simulation results for the whole range of $\alpha$ is observed. This means that the molecular chaos ansatz does not limit the applicability of the Boltzmann description, even for large inelasticities~\cite{G19}, at~least for dilute granular gases. As~for the approximate theoretical predictions, it is quite apparent that $a_2^{\hcs,b}$ (see Equation \eqref{a2b}) performs very well, even if the fourth cumulant is not small (e.g., $a_2^{\hcs}$$\sim$0.2 at $\alpha=0.1$). The~approximate sixth cumulant $a_3^{\hcs,ab}$ (see Equations \eqref{n10} and \eqref{a3ab}) is less accurate at a quantitative level, especially in the case of disks, but~captures quite well the general influence of inelasticity. While $a_2^\hcs$ changes from negative to positive values at $\alpha\simeq 1/\sqrt{2}\simeq 0.71$, $a_3^\hcs$ is always negative.
Note that, for~large inelasticity, the~cumulants $a_2^\hcs$ and $a_3^\hcs$ are comparable in magnitude. Given that the Sonine expansion \eqref{eq:fv} is only asymptotic~\cite{BP04,NBSG07}, it is remarkable that a theoretical approach based on the assumptions $|a_3^\hcs|\ll |a_2^\hcs|\ll 1$ does such a good job for high inelasticity as observed in Figure~\ref{fig:a2a3st}.

Next, we study the evolution from a non-HCS state, as~monitored by $a_2(s)$ and $a_3(s)$. We have chosen an initial state very far from the HCS: the particles are arranged in an ordered crystalized configuration and all have a common speed $\sqrt{d/2}v_\thr(0)$ along uniformly randomized directions. Therefore, at~$s=0$, $\braket{c^k}=(d/2)^{k/2}$, so that $a_2(0)=-\frac{2}{d+2}$ and $a_3(0)=-\frac{16}{(d+2)(d+4)}$.

\begin{figure}[H]
\centering
\includegraphics[width=.45\columnwidth]{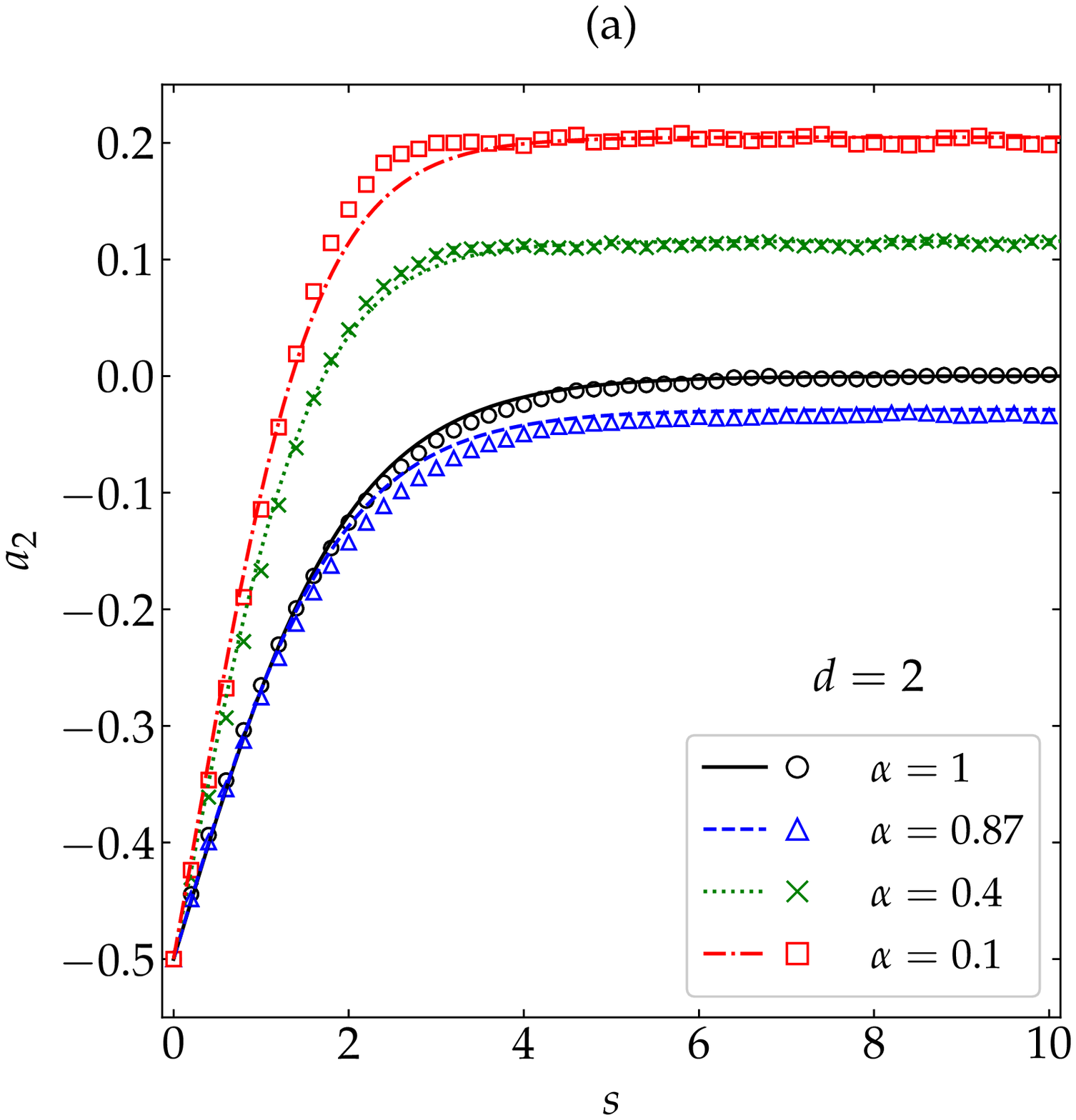}\hspace{1cm}\includegraphics[width=.45\columnwidth]{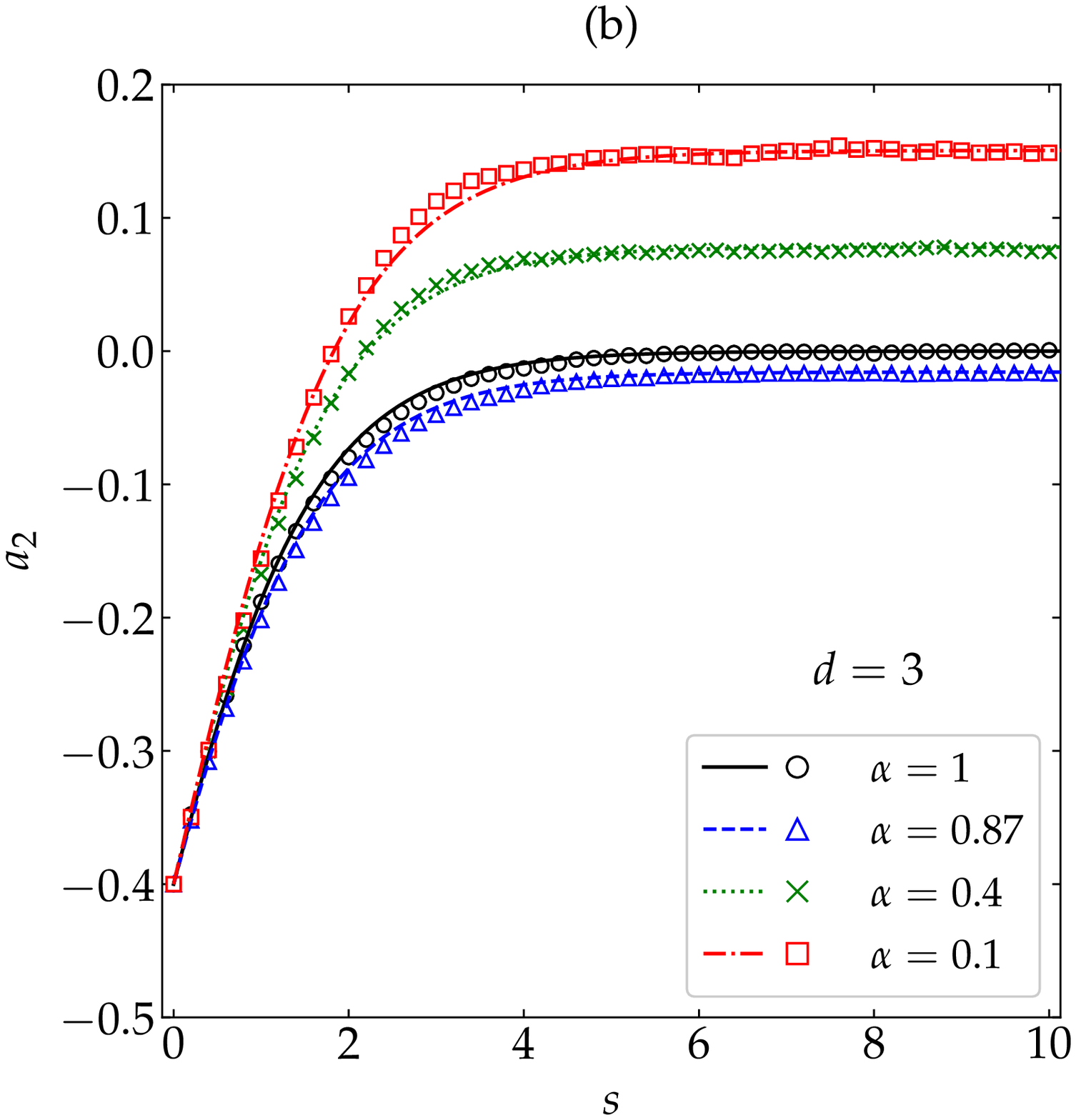}
\caption{Evolution of the fourth cumulant $a_2(s)$ as a function of the average number of collisions per particle for (\textbf{a}) disks and (\textbf{b}) spheres. Symbols represent MD simulation results, while the lines correspond to the theoretical prediction \eqref{Sol_a2}. The~values of the coefficient of restitution are (from top to bottom) $\alpha=0.1$ ($\square$), $0.4$ ($\times$), $1$ ($\circ$), and~$0.87$ ($\triangle$). The~error bars in the simulation data are smaller than the size of the~symbols.\label{fig:a2Evol}}
\end{figure}

\begin{figure}[H]
\centering
\includegraphics[width=0.45\columnwidth]{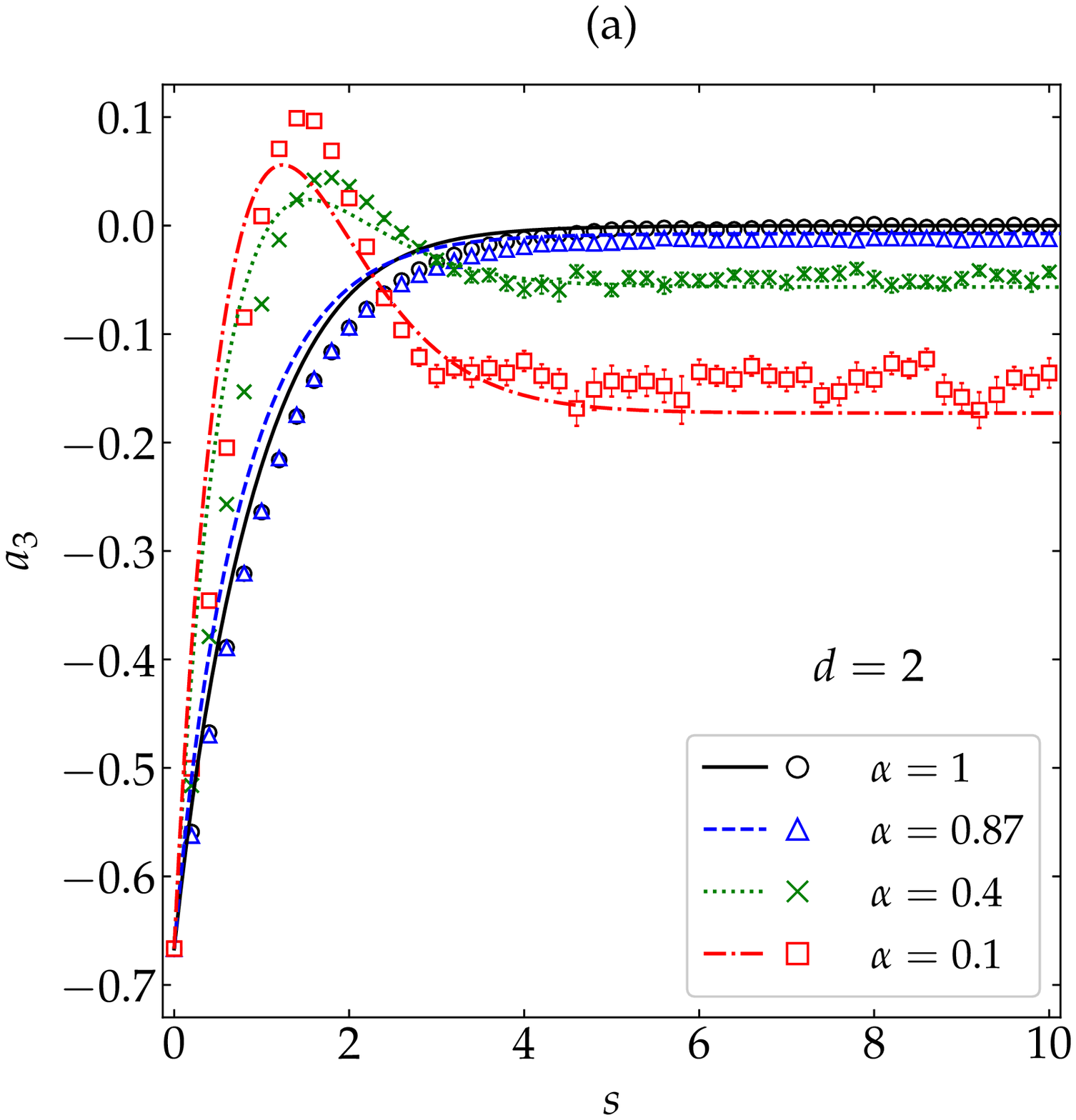}\hspace{1cm}\includegraphics[width=.45\columnwidth]{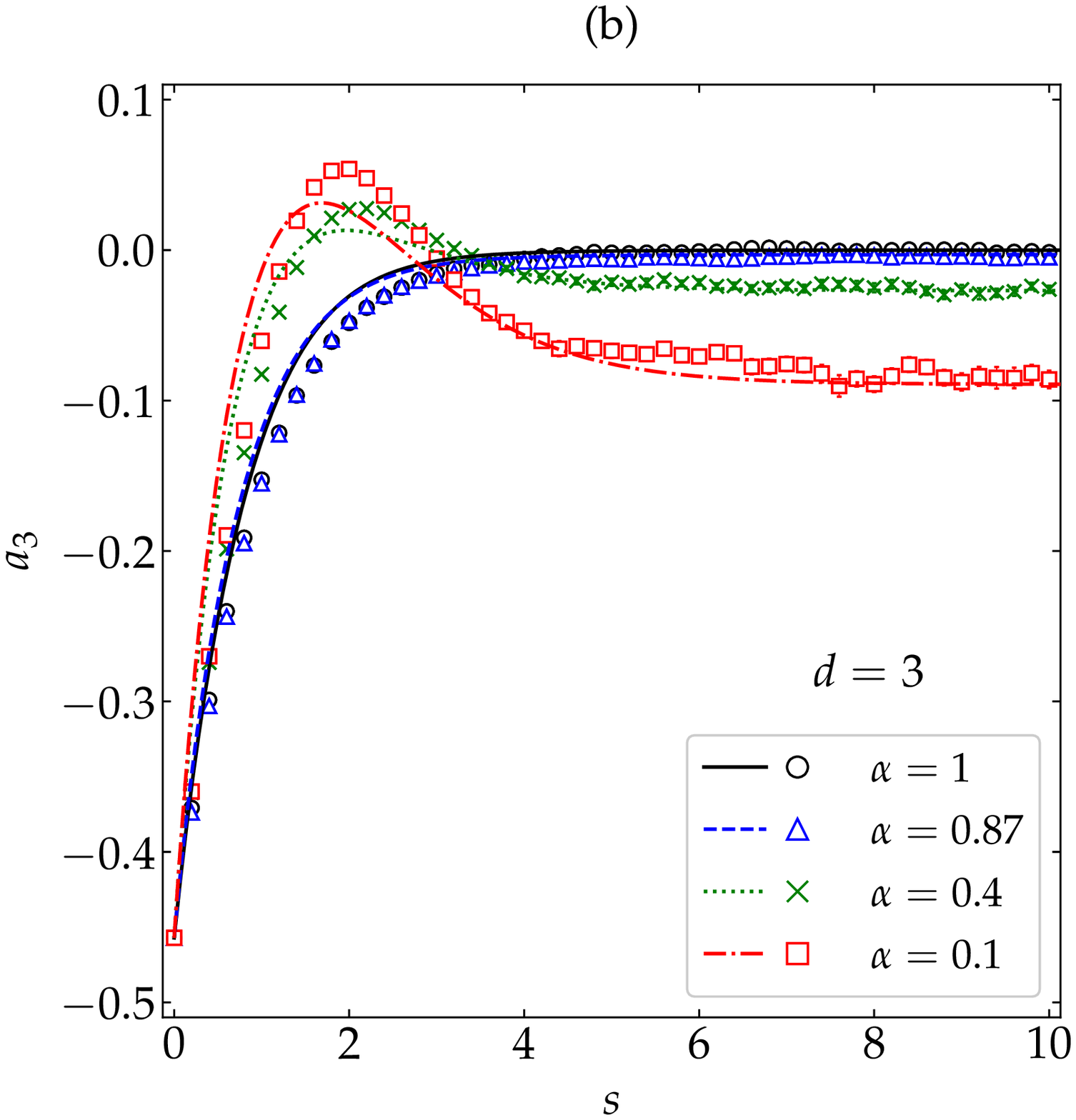}
\caption{ {Evolution} %Authors: Size of figures increased. Please use always either width=0.45\columnwidth or width=0.35\columnwidth
of the sixth cumulant $a_3(s)$ as a function of the average number of collisions per particle for (\textbf{a}) disks and (\textbf{b}) spheres. Symbols represent MD simulation results, while the lines correspond to the theoretical prediction \eqref{Sol_a3}. The~values of the coefficient of restitution are (from~bottom to top on the right side) $\alpha=0.1$ ($\square$), $0.4$ ($\times$), $0.87$ ($\triangle$), and~$1$ ($\circ$). The~error bars in the simulation data are smaller than the size of the symbols, except~in the stationary regime for $\alpha=0.1$.\label{fig:a3Evol}}
\end{figure}

Figures~\ref{fig:a2Evol} and \ref{fig:a3Evol} compare our MD results with the theoretical predictions \eqref{Sol_a2} and \eqref{Sol_a3}, respectively. Four representative values of the coefficient of restitution have been considered, namely $\alpha=0.1$ (very~high inelasticity), $0.4$ (high inelasticity), $0.87$ (moderately small inelasticity), and~$1$ (elastic~collisions); $\alpha=0.87$ has been included because it is practically at this value where $a_2^\hcs$ presents a local minimum, both for disks and spheres [see Figure \ref{fig:a2a3st}a]. Note that, in~the case of simulations, the~quantity $s$  represents the \emph{actual} average number of collisions per particle and, consequently, is not strictly defined by Equation \eqref{s(t)}, in~contrast to the case of theory.
From Figure~\ref{fig:a2Evol} we observe that, despite the large magnitude of the initial fourth cumulant ($a_2(0)=-\frac{1}{2}$ and $-\frac{2}{5}$ for $d=2$ and $3$, respectively), the~simple relaxation law \eqref{Sol_a2} describes very well the full evolution of the cumulant. Discrepancies with the simulation results are visible only in the region ($2\lesssim s \lesssim 4$) where the curves turn to their stationary values, especially in the case of disks. In~what concerns the sixth cumulant, which also has a large initial magnitude ($a_3(0)=-\frac{2}{3}$ and $-\frac{16}{35}$ for $d=2$ and $3$, respectively), the~theoretical expression \eqref{Sol_a3} is able to capture, at~least, the~main qualitative features, including the change from a non-monotonic ($\alpha=0.1$ and $0.4$) to a monotonic ($\alpha=0.87$ and $1$) evolution. Again, the~agreement is better for spheres than for disks. Note also that the evolution curves for $\alpha=0.87$ and $1$ are hardly distinguishable from each~other.
\unskip

\unskip

\begin{figure}[H]
\centering
\includegraphics[width=.45\columnwidth]{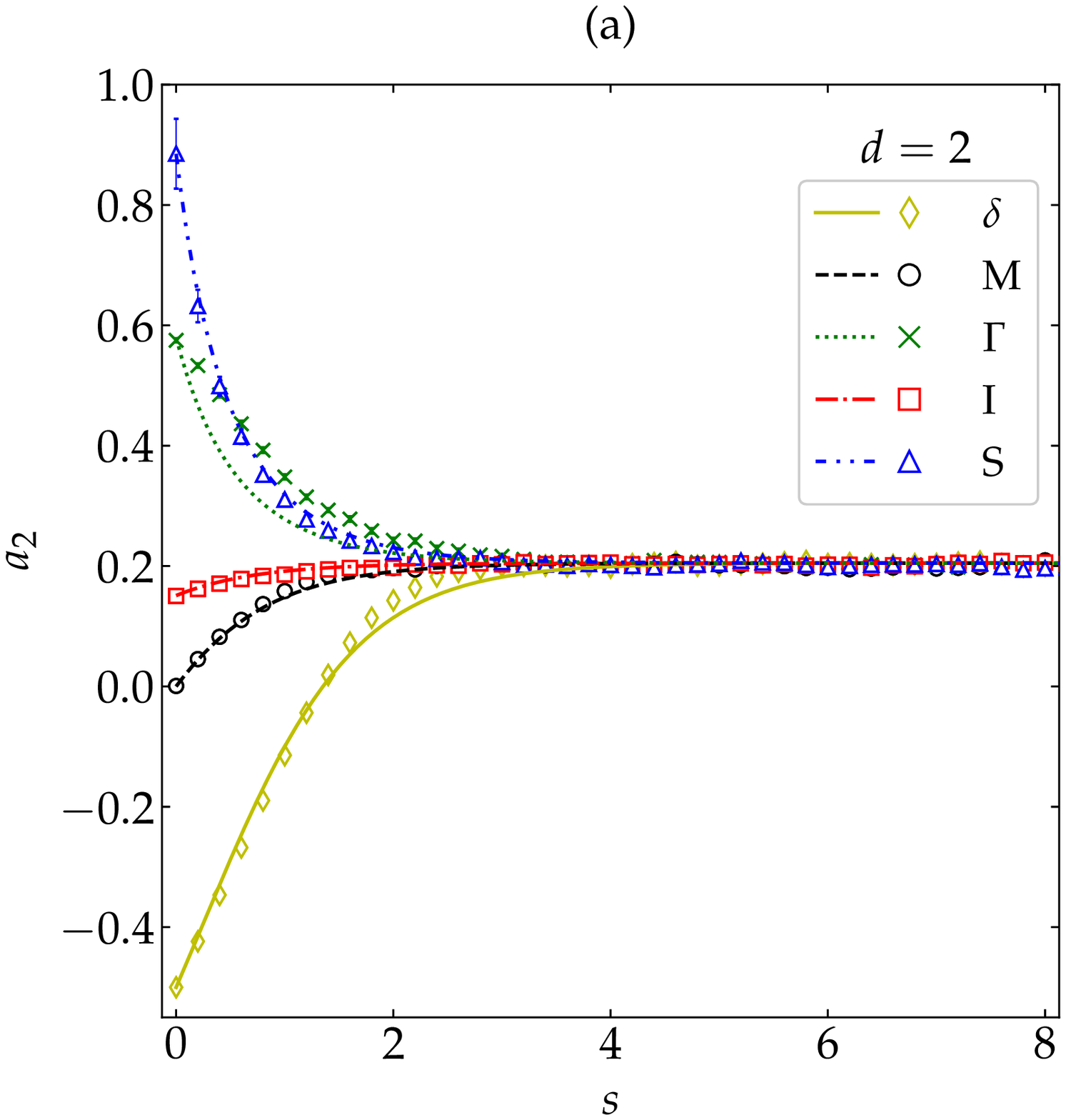}\hspace{1cm}\includegraphics[width=.45\columnwidth]{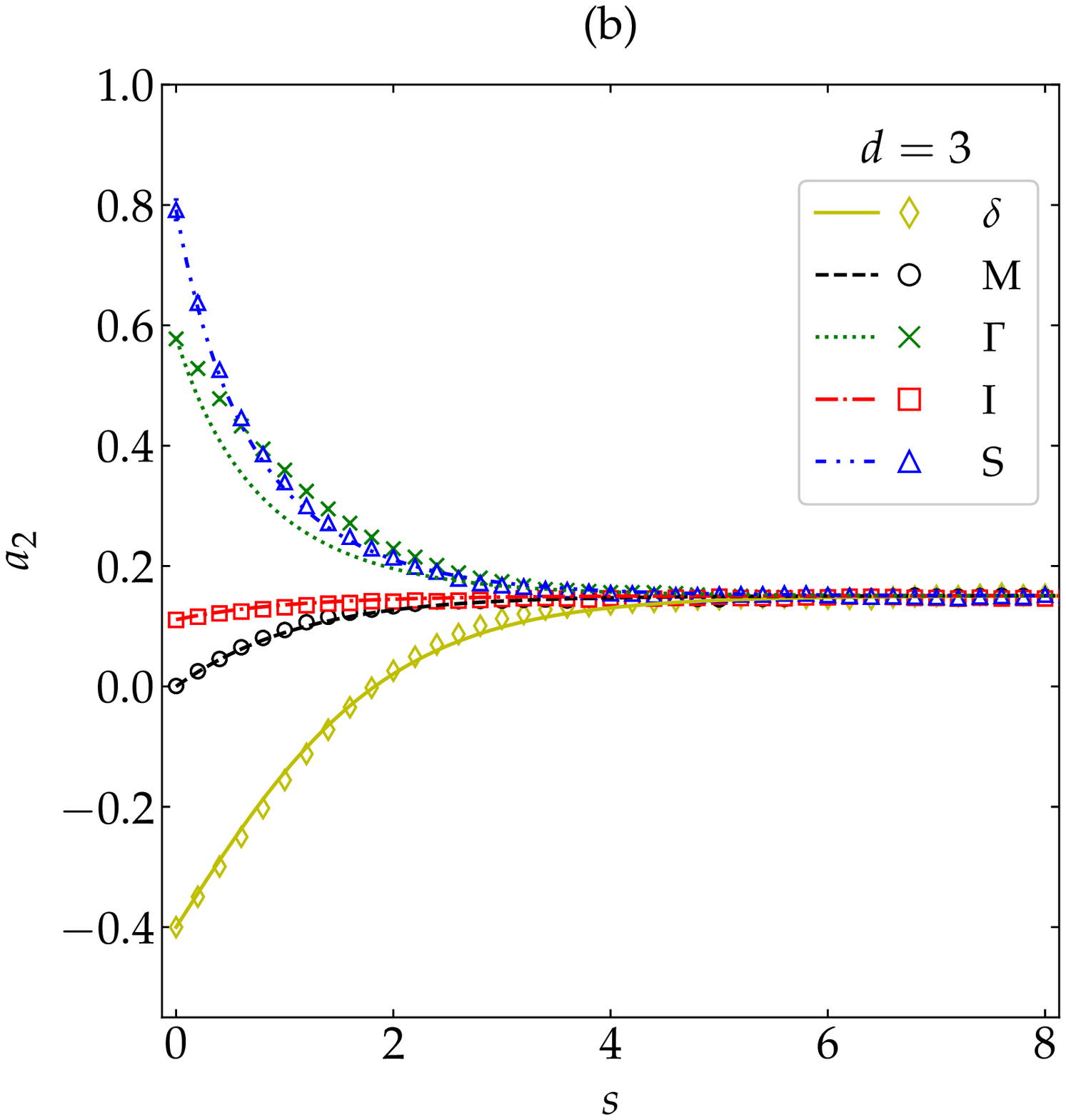}
\caption{ {Evolution} %Authors: Size of figures increased. Please use always either width=0.45\columnwidth or width=0.35\columnwidth
of the fourth cumulant $a_2(s)$ for a coefficient of restitution  $\alpha=0.1$ as a function of the average number of collisions per particle for (\textbf{a}) disks and (\textbf{b}) spheres. Symbols represent MD simulation results, while the lines correspond to the theoretical prediction \eqref{Sol_a2}. Five different initial conditions are considered (see Appendix \ref{app:C}): $\delta$ ($\diamond$), M ($\circ$), $\Gamma$ ($\times$), I ($\square$), and~S ($\triangle$). The~error bars in the simulation data are smaller than the size of the symbols, except~in the early stage for the initial condition~S.\label{fig:a2Evol_new}}
\end{figure}

\begin{figure}[H]
\centering
\includegraphics[width=.45\columnwidth]{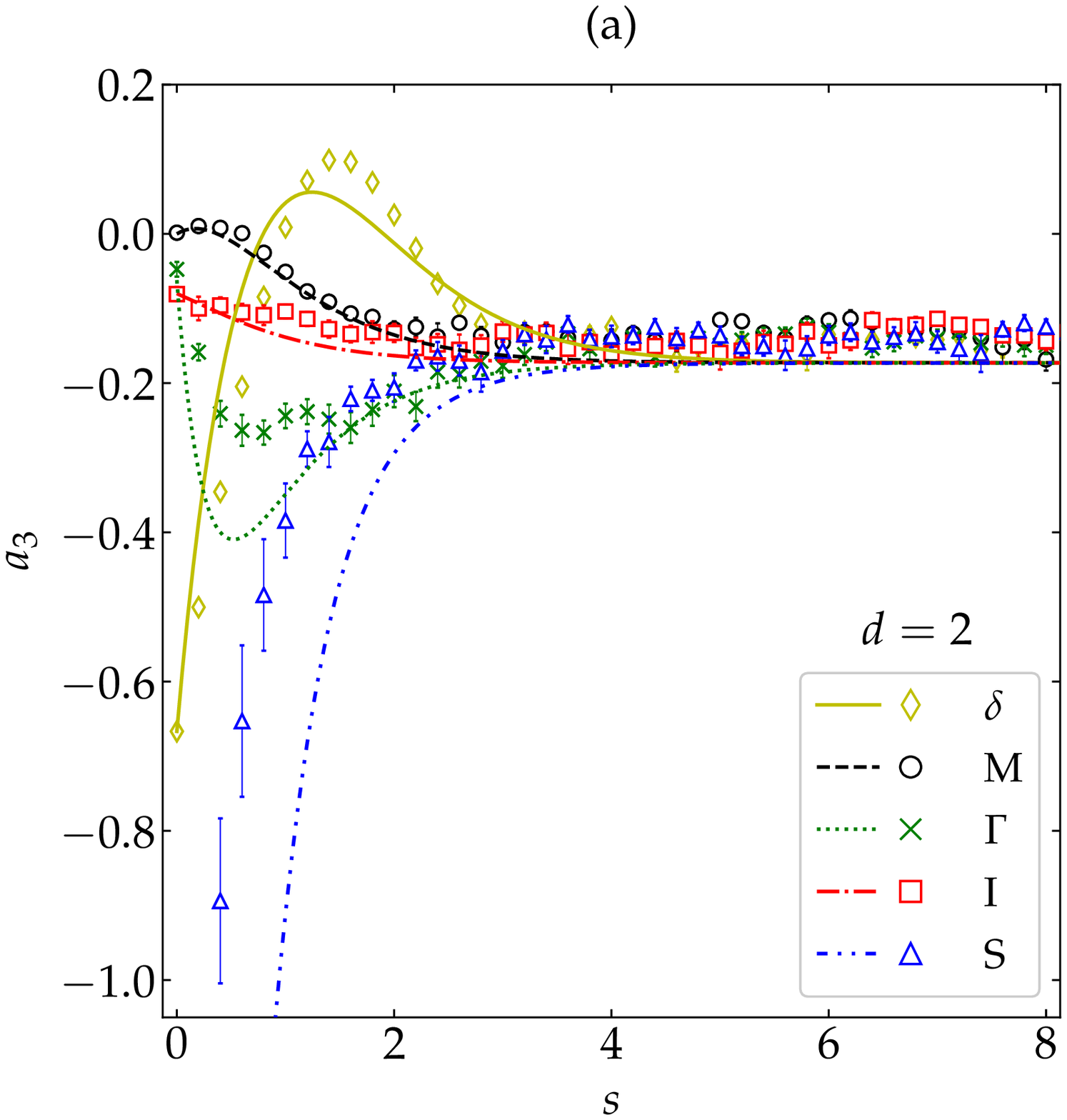}\hspace{.9cm}\includegraphics[width=.45\columnwidth]{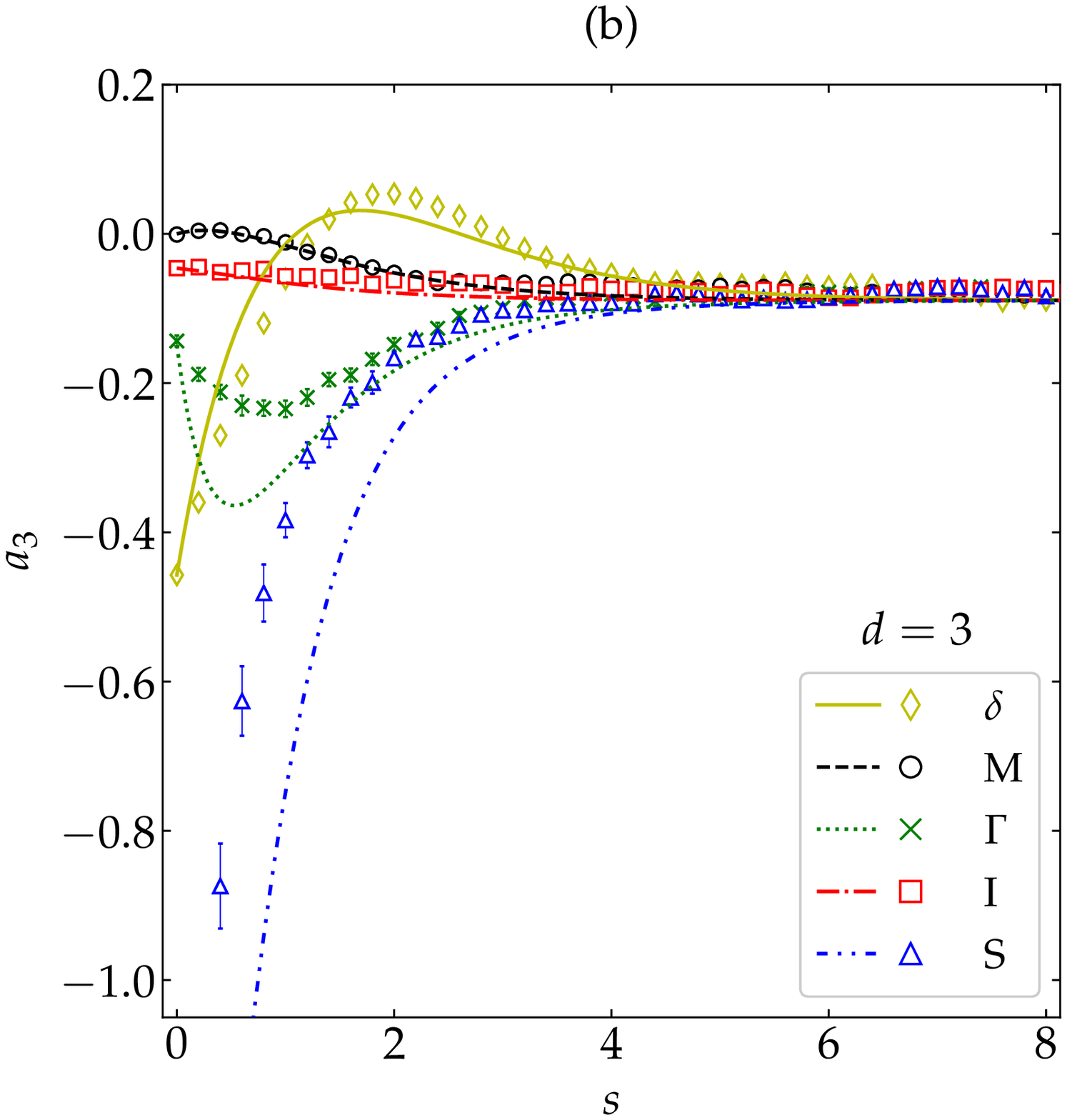}
\caption{Evolution of the sixth cumulant $a_3(s)$ for a coefficient of restitution  $\alpha=0.1$ as a function of the average number of collisions per particle for (\textbf{a}) disks and (\textbf{b}) spheres. Symbols represent MD simulation results, while the lines correspond to the theoretical prediction \eqref{Sol_a3}. Five different initial conditions are considered (see Appendix \ref{app:C}): $\delta$ ($\diamond$), M ($\circ$), $\Gamma$ ($\times$), I ($\square$), and~S ($\triangle$). The~error bars in the simulation data are smaller than the size of the symbols, except~in the early stage for the initial condition~S.\label{fig:a3Evol_new}}
\end{figure}

In Figures~\ref{fig:a2Evol} and \ref{fig:a3Evol}, the~initial values $a_2(0)$ and $a_3(0)$ are common to all the coefficients of restitution considered. In~order to have a more complete picture, let us now fix the most inelastic systems  ($\alpha=0.1$) and take five different initial conditions. The~HCS values of the fourth and sixth cumulants at $\alpha=0.1$ are $\{a_2^\hcs,a_3^\hcs\}=\{0.206,-0.143\}$ and $\{0.150,-0.077\}$ for $d=2$ and $d=3$, respectively. Thus, we have chosen the same initial distribution (hereafter labeled as $\delta$) as in Figures~\ref{fig:a2Evol} and \ref{fig:a3Evol} as a representative example of $a_2(0)<0$, the~Maxwellian distribution (labeled as M) with $a_2(0)=0$, another one (\mbox{labeled as I}) with $0<a_2(0)<a_2^\hcs$, and~two more (labeled as $\Gamma$ and S) with $a_2(0)>a_2^\hcs$. The~details of those five distributions can be found in Appendix \ref{app:C} and the corresponding values of $a_2(0)$ and $a_3(0)$ are shown in Table~\ref{table0}.
 In the case of $a_2(s)$, Figure~\ref{fig:a2Evol_new} shows again an excellent agreement between theory and simulation, except~for the initial condition $\Gamma$ and near the turning point already observed in Figure~\ref{fig:a2Evol} for the initial condition $\delta$. In~what concerns $a_3(s)$, one can observe from Figure~\ref{fig:a3Evol_new} that the performance of the approximation \eqref{Sol_a3} is generally fair, especially for the initial conditions M and I. The~limitations of Equation \eqref{Sol_a2} for the initial condition $\Gamma$ and of Equation \eqref{Sol_a3} for the initial conditions $\Gamma$, S, and~$\delta$ are due to the role played by higher-order cumulants in those~cases.

\unskip

%%%%%%%%%%%%%%%%%%%%%%%%%%%%%%%%%%%%%%%%%%
\section{KLD as a Lyapunov~Functional}
\label{sec3}
 {In this section}%Authors: Parentheses removed and sentence moved
, we restrict ourselves to spatially homogeneous states.
\vspace{-6pt}
\subsection{Boltzmann's $H$-Functional}

The introduction of the $H$-theorem by Ludwig Boltzmann~\cite{B95} was a revolution in physics and became an inspiration for new mathematical and physical concepts. This theorem is a direct consequence of the Boltzmann kinetic equation for classical rarefied gases, derived under its molecular chaos assumption~\cite{CC70,GS03}. Beneath this hypothesis for a classical gas which evolves via \emph{elastic} collisions, the~$H$-functional
{defined as}
\begin{equation}\label{eq:H_def}
    H(t) = \int \dif\vv f(\vv;t)\ln f(\vv;t)
\end{equation}
is proved to be a non-increasing quantity; in other words, $S=-H$, up~to a constant, is a non-decreasing entropy-like Lyapunov functional for the assumed gaseous system. After~almost a century, once Information Theory was developed, Boltzmann's $H$-functional was interpreted as Shannon's measure~\cite{S48} for the one-particle  VDF  of a rarefied~gas.

Nonetheless, the~model considered in this paper for a rarefied monocomponent granular gas (inelastic and smooth hard $d$-spheres with a constant coefficient of restitution) violates Boltzmann's hypothesis of elastic collisions. In~fact, a~key role in the demonstration of the $H$-theorem for elastic collisions is played by the condition of
 {collisional symmetry} %Authors: Italics removed
 \cite{B95}. Consider two colliding particles with precollision velocities $\{\vv_1'',\vv_2''\}$ and a relative orientation characterized by the unit vector $-\s$ (with~$\vv_{12}''\cdot\s<0$). After~collision, the~velocities are, in~agreement with Equation \eqref{coll_rule}, given by
\begin{equation}
\label{C31}
\mathfrak{C}_{-\s}\{\vv_1'',\vv_2''\}=\{\vv_1,\vv_2\},\quad \vv_{1,2}=\vv_{1,2}''\mp\frac{1+\alpha}{2}(\vv_{12}''\cdot\s)\s.
\end{equation}
Next, suppose two colliding particles with precollision velocities $\{\vv_1,\vv_2\}$ and a relative orientation characterized by the unit vector $\s$ (with $\vv_{12}\cdot\s>0$). In~that case,
\begin{equation}
\label{C32}
\mathfrak{C}_{\s}\{\vv_1,\vv_2\}=\mathfrak{C}_{\s}\mathfrak{C}_{-\s}\{\vv_1'',\vv_2''\}=\{\vv_1',\vv_2'\},\quad \vv_{1,2}'=\vv_{1,2}\mp\frac{1+\alpha}{2}(\vv_{12}\cdot\s)\s.
\end{equation}
Comparison with Equation \eqref{coll_rule} shows that
\begin{equation}
\vv_{1,2}'=\vv_{1,2}''\pm\frac{1-\alpha^2}{2\alpha}(\vv_{12}\cdot\s)\s,\quad \vv_{12}'\cdot\s=\alpha^2\vv_{12}''\cdot\s.
\end{equation}
Thus, $\mathfrak{C}_{\s}\mathfrak{C}_{-\s}\{\vv_1'',\vv_2''\}\neq \{\vv_1'',\vv_2''\}$ unless $\alpha=1$ and, therefore, the~$H$-functional, as~defined by Equation~\eqref{eq:H_def}, is not ensured to be non-increasing anymore if $\alpha<1$.

Furthermore, Boltzmann's $H$-functional for the model of inelastic particles presents the so-called measure problem~\cite{MT11}. Shannon's measure is invariant under unitary transformations, but~not for rescaling. In~fact, under~the transformation \eqref{eq:phi},
\begin{equation}
\label{H*}
    H(s) = \int \dif\vv\medspace f(\vv,t)\ln f(\vv,t)=H^*(s)  -\frac{d}{2}\ln \frac{2T(s)}{m},\quad H^*(s)\equiv \int \dif\cc \medspace \phi(\cc,s)\ln \phi(\cc,s).
\end{equation}
From Haff's law, Equation \eqref{eq:Haff}, it turns out that (in the HCS) $H_\hcs^*$ is  stationary but $H_\hcs(s)$ grows linearly with the average number of collisions $s$. Then, one could naively think that a possible candidate to the Lyapunov  functional would be $H^*(s)$, but~the latter is still non-invariant under a change of variables $\cc\to \widetilde{\cc}=\bm{w}(\cc)$, $\phi(\cc,s)\to\widetilde{\phi}(\widetilde{\cc},s)=J^{-1}\phi(\cc,s)$, where $J\equiv |\partial\widetilde{\cc}/\partial \cc|$ is the Jacobian of the invertible transformation  $\widetilde{\cc}=\bm{w}(\cc)$.
As will be seen below, whereas Shannon's measure presents a problematic  weighting of the phase space, the~KLD solves this non-invariance~issue.

\subsection{KLD}

In general, given two distribution functions $f(\bm{x})$ and $g(\bm{x})$, one  defines the KLD from $g$ to $f$ (or~\emph{relative entropy} of $f$ with respect to $g$) as~\cite{KL51,K78}, as~\begin{equation}
    \mathcal{D}_{\text{KL}}(f\|g)=\int_{X} \dif \bm{x}\medspace f(\bm{x})\ln\frac{f(\bm{x})}{g(\bm{x})},
\end{equation}
where $\bm{x}$ is  a random vector  variable defined on the set $X$. The~quantity $\mathcal{D}_{\text{KL}}(f\|g)$ is convex and non-negative, being identically zero if and only if $f=g$. While it is not a distance or metric function (it does not obey either symmetry or triangle inequality properties), $\mathcal{D}_{\text{KL}}(f\|g)$ somehow measures how much a \emph{reference} distribution $g$ \emph{diverges} from the actual distribution $f$ or, equivalently, the~amount of information lost when $g$ is used to approximate $f$.

Therefore, it seems convenient to define the KLD
\begin{equation}
\label{KLD}
\mathcal{D}_{\text{KL}}(f\|f_\refe)=\mathcal{D}_{\text{KL}}(\phi\|\phi_\refe)=\int \dif\cc\medspace  \phi(\cc;s)\ln\frac{\phi(\cc;s)}{\phi_\refe(\cc)}
\end{equation}
as the entropy-like Lyapunov functional for our problem, where the (stationary) reference function $\phi_\refe$ must be an \emph{attractor} to ensure the Lyapunov-functional condition.
Thus, if~we choose \mbox{$\phi_\refe(\cc)=\lim_{s\rightarrow\infty} \phi(\cc;s)$}, assuming that this limit exists, it will minimize the KLD for asymptotically long times.  In~addition, the~definition \eqref{KLD} solves the measure problem  posed above, i.e.,~\mbox{$\mathcal{D}_{\text{KL}}(\phi\|\phi_\refe)=\mathcal{D}_{\text{KL}}(\widetilde{\phi}\|\widetilde{\phi}_\refe)$} for any invertible transformation  $\cc\to \widetilde{\cc}=\bm{w}(\cc)$.

If $\mathcal{D}_{\text{KL}}(\phi\|\phi_\refe)$ is indeed the Lyapunov functional of our problem, the~natural conjecture is that $\phi_\refe(\cc)=\phi_\hcs(\cc)$ \cite{GMMMRT15}. As~a consequence, the~challenge is to prove that $\partial_s\mathcal{D}_{\text{KL}}(\phi\|\phi_\hcs)\leq 0$ (see Appendix~\ref{app:A0} for a formal expression of $\partial_s\mathcal{D}_{\text{KL}}(\phi\|\phi_\refe)$ in the context of the inelastic Boltzmann equation). While in this paper we do not intend to address such a proof from a mathematical point of view, we will provide support by means of MD simulations (see Appendix \ref{app:A} for technical details). Before~doing that, and~in order to put the problem in a proper context, we consider the alternative choice $\phi_\refe=\phi_{\text{M}}$.

\subsection{MD~Simulations}
\unskip
\subsubsection{Maxwellian Distribution as a Reference ($\phi_\refe=\phi_{\text{M}}$)}
If $\phi_\refe=\phi_{\text{M}}$ is chosen in Equation \eqref{KLD}, one simply has
\begin{equation}
\mathcal{D}_{\text{KL}}(\phi\|\phi_{\text{M}})=H^*(s)+\frac{d}{2}\left(1+\ln\pi\right),
\end{equation}
where $H^*(s)$ is defined in Equation \eqref{H*}. Thus, $\mathcal{D}_{\text{KL}}(\phi\|\phi_{\text{M}})$ differs from $H^*(s)$ by a constant, so that $\partial_s\mathcal{D}_{\text{KL}}(\phi\|\phi_{\text{M}})=\partial_s H^*(s)$.

Note that $\partial_s\mathcal{D}_{\text{KL}}(\phi\|\phi_{\text{M}})$ cannot be semi-definite negative for \emph{arbitrary} initial conditions. For~instance, if~the initial condition is a Maxwellian, i.e.,~$\phi(\cc;0)=\phi_{\text{M}}(\cc)$, then it is obvious that $\left.\mathcal{D}_{\text{KL}}(\phi\|\phi_{\text{M}})\right|_{s=0}=0$ and, given that $\lim_{s\to\infty}\mathcal{D}_{\text{KL}}(\phi\|\phi_{\text{M}})=\mathcal{D}_{\text{KL}}(\phi_\hcs\|\phi_{\text{M}})>0$, it is impossible that $\partial_s\mathcal{D}_{\text{KL}}(\phi\|\phi_{\text{M}})\leq 0$ for all $s$. Nevertheless, in~principle, it might happen that  $\partial_s\mathcal{D}_{\text{KL}}(\phi\|\phi_{\text{M}})\leq 0$ for the class of initial conditions such that $\left.\mathcal{D}_{\text{KL}}(\phi\|\phi_{\text{M}})\right|_{s=0}\geq \mathcal{D}_{\text{KL}}(\phi_\hcs\|\phi_{\text{M}})$, while $\partial_s\mathcal{D}_{\text{KL}}(\phi\|\phi_{\text{M}})\geq 0$ for the complementary class of initial conditions such that $\left.\mathcal{D}_{\text{KL}}(\phi\|\phi_{\text{M}})\right|_{s=0}\leq \mathcal{D}_{\text{KL}}(\phi_\hcs\|\phi_{\text{M}})$. If~that were the case, one could say that the quantity $\left[\mathcal{D}_{\text{KL}}(\phi\|\phi_{\text{M}})-\mathcal{D}_{\text{KL}}(\phi_\hcs\|\phi_{\text{M}})\right]^2$ would always decrease for every initial condition, thus qualifying as a Lyapunov functional. As~will be seen below, this expectation is frustrated by our simulation~results.

From the formal Sonine expansion \eqref{eq:fv}, one has
\begin{equation}
\mathcal{D}_{\text{KL}}(\phi\|\phi_{\text{M}})=\int\dif\cc\,\phi_{\text{M}}(\cc)\left[1+\sum_{k=2}^\infty a_k(s)S_k(c^2)\right]\ln \left[1+\sum_{k=2}^\infty a_k(s)S_k(c^2)\right].
\end{equation}
Now, in~the spirit of the truncation approximation of Section~\ref{sec2.3}, we can write the approximate~expression
\begin{equation}
\label{DKLM}
\mathcal{D}_{\text{KL}}(\phi\|\phi_{\text{M}})\approx\int\dif\cc\,\phi_{\text{M}}(\cc)\left[1+ a_2(s)S_2(c^2)+ a_3(s)S_3(c^2)\right]\ln \left[1+ a_2(s)S_2(c^2)+ a_3(s)S_3(c^2)\right],
\end{equation}
where $a_2(s)$ and $a_3(s)$ are given by Equations \eqref{Sol_a2} and \eqref{Sol_a3}, respectively.
Since the truncated Sonine approximation is not positive definite, we will take the real part of the right-hand side of Equation \eqref{DKLM} for times such that $1+ a_2(s)S_2(c^2)+ a_3(s)S_3(c^2)<0$ for a certain range of~velocities.

\begin{figure}[H]
\centering
\includegraphics[width=.45\columnwidth]{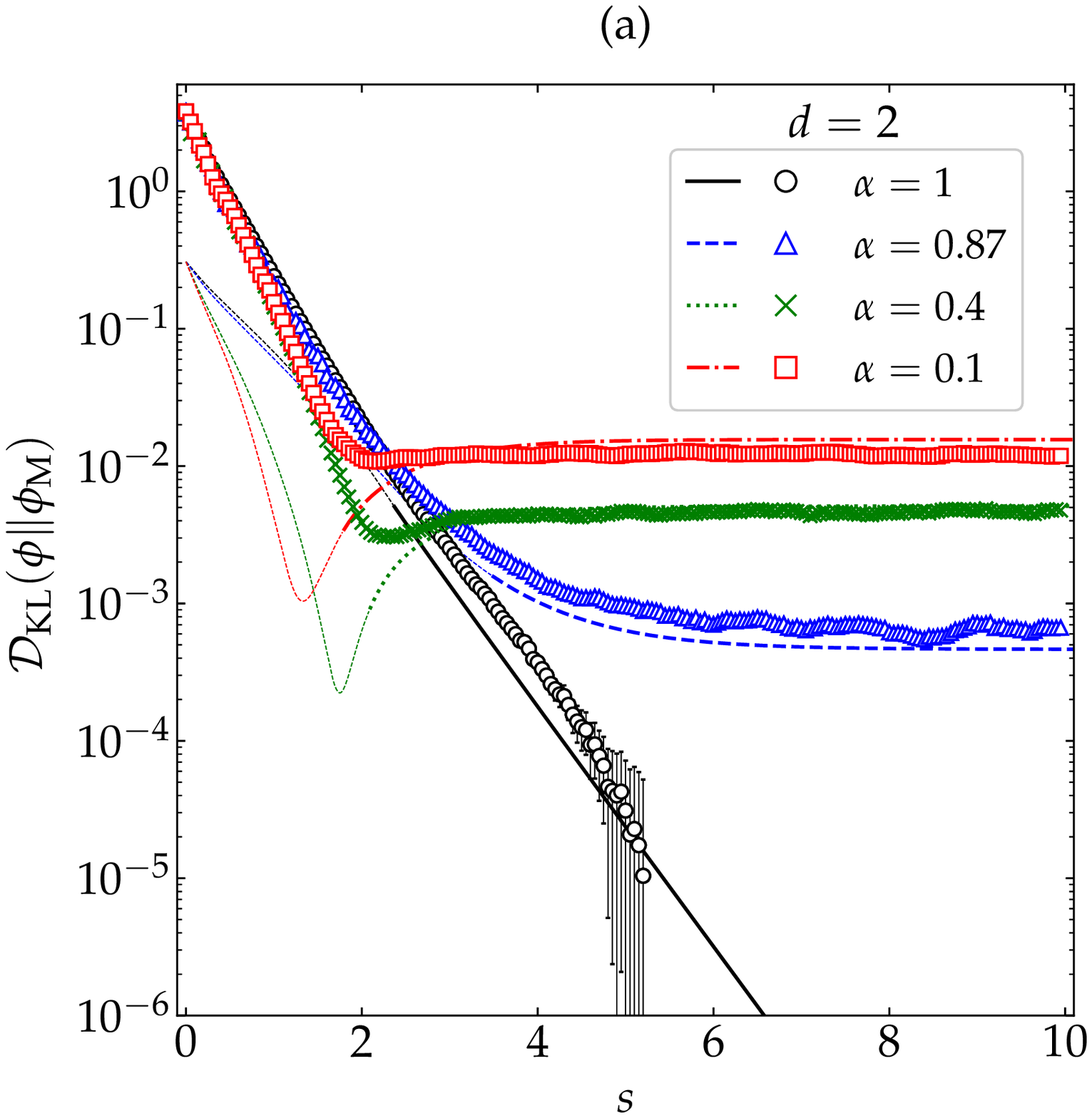}\hspace{.9cm}\includegraphics[width=.45\columnwidth]{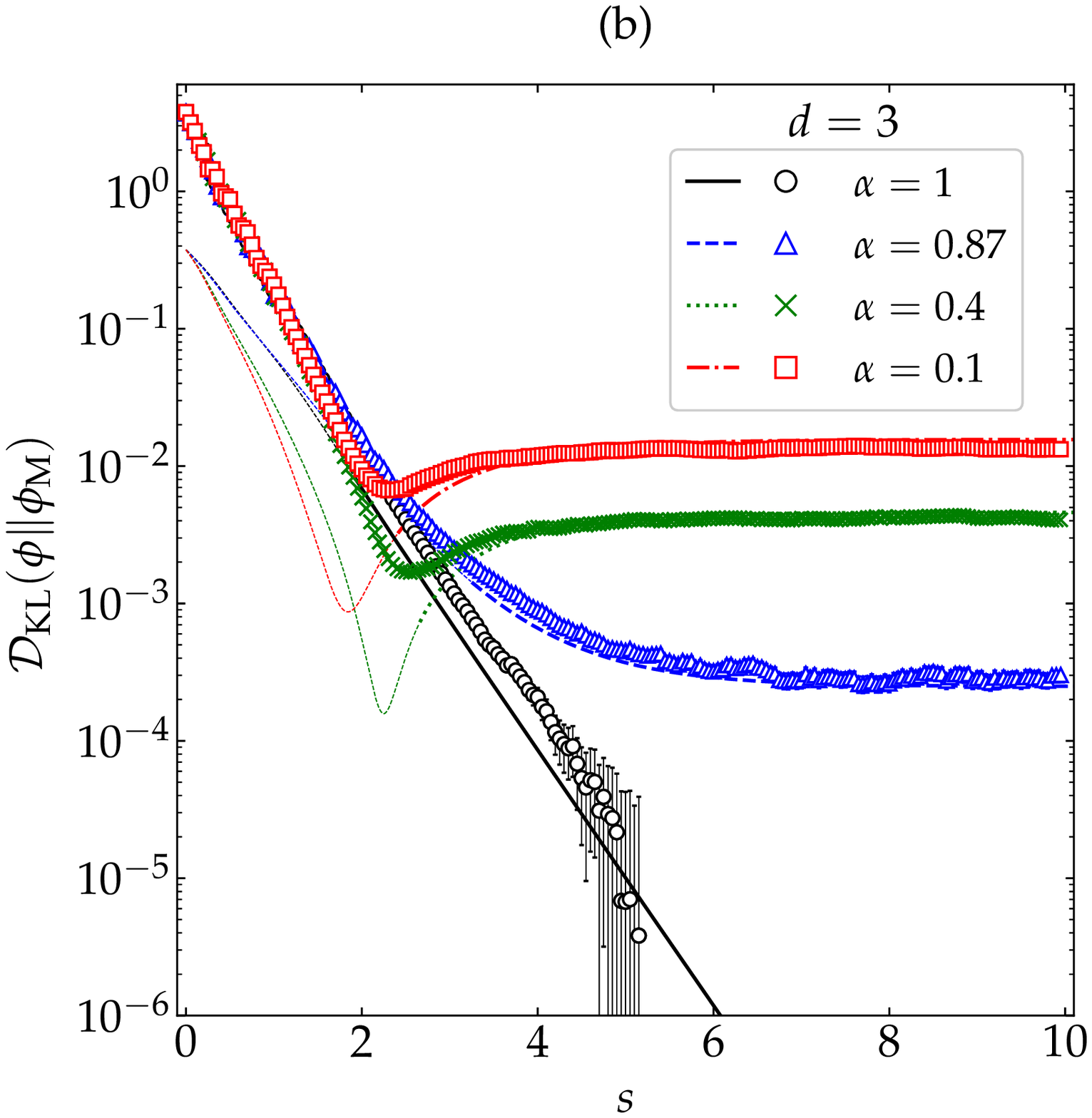}
\caption{Evolution of  $\mathcal{D}_{\text{KL}}(\phi\|\phi_{\text{M}})$ (in logarithmic scale) as a function of the average number of collisions per particle for (\textbf{a}) disks and (\textbf{b}) spheres. Symbols represent MD simulation results, while the lines correspond to the theoretical approximation \eqref{DKLM} (the thin dashed lines for the first stage of the evolution mean that it was necessary to take the real part). The~values of the  coefficient of restitution are (from top to bottom on the right side)  $\alpha=0.1$ ($\square$), $0.4$ ($\times$), $0.87$ ($\triangle$), and~$1$ ($\circ$). The~error bars in the simulation data are smaller than the size of the symbols, except~when $\mathcal{D}_{\text{KL}}(\phi\|\phi_{\text{M}})\lesssim 10^{-4}$ for $\alpha=1$. \label{fig:DKL_M_evl}}
\end{figure}

Figure~\ref{fig:DKL_M_evl} shows the evolution of  $\mathcal{D}_{\text{KL}}(\phi\|\phi_{\text{M}})$ for the same initial conditions and the same values of $\alpha$ as in Figures~\ref{fig:a2Evol} and \ref{fig:a3Evol}, as~obtained from our MD simulations (for details, see Appendix \ref{app:A}) and from the crude approximation \eqref{DKLM}. For~that initial condition, one clearly has
\mbox{$\left.\mathcal{D}_{\text{KL}}(\phi\|\phi_{\text{M}})\right|_{s=0}>\mathcal{D}_{\text{KL}}(\phi_\hcs\|\phi_{\text{M}})$}. A~monotonic behavior
$\partial_s\mathcal{D}_{\text{KL}}(\phi\|\phi_{\text{M}})\leq 0$ is observed only in the cases of small or vanishing inelasticity. For~$\alpha=0.1$ and $0.4$, however, $\mathcal{D}_{\text{KL}}(\phi\|\phi_{\text{M}})$ does not present a monotonic decay but tends to its asymptotic value $\mathcal{D}_{\text{KL}}(\phi_\hcs\|\phi_{\text{M}})$ from below, there existing a time ($s$$\sim$2) at which  $\mathcal{D}_{\text{KL}}(\phi\|\phi_{\text{M}})$ exhibits a local minimum. This non-monotonic behavior is certainly exaggerated by the truncated Sonine approximation \eqref{DKLM}, but~it is clearly confirmed by our MD simulations, especially in the case of spheres.
Therefore, it is quite obvious that, not unexpectedly, both $\mathcal{D}_{\text{KL}}(\phi\|\phi_{\text{M}})$ and $\left[\mathcal{D}_{\text{KL}}(\phi\|\phi_{\text{M}})-\mathcal{D}_{\text{KL}}(\phi_\hcs\|\phi_{\text{M}})\right]^2$ must be discarded as a Lyapunov functional for the free cooling of granular~gases.

\begin{figure}[H]
\centering
\includegraphics[width=.45\columnwidth]{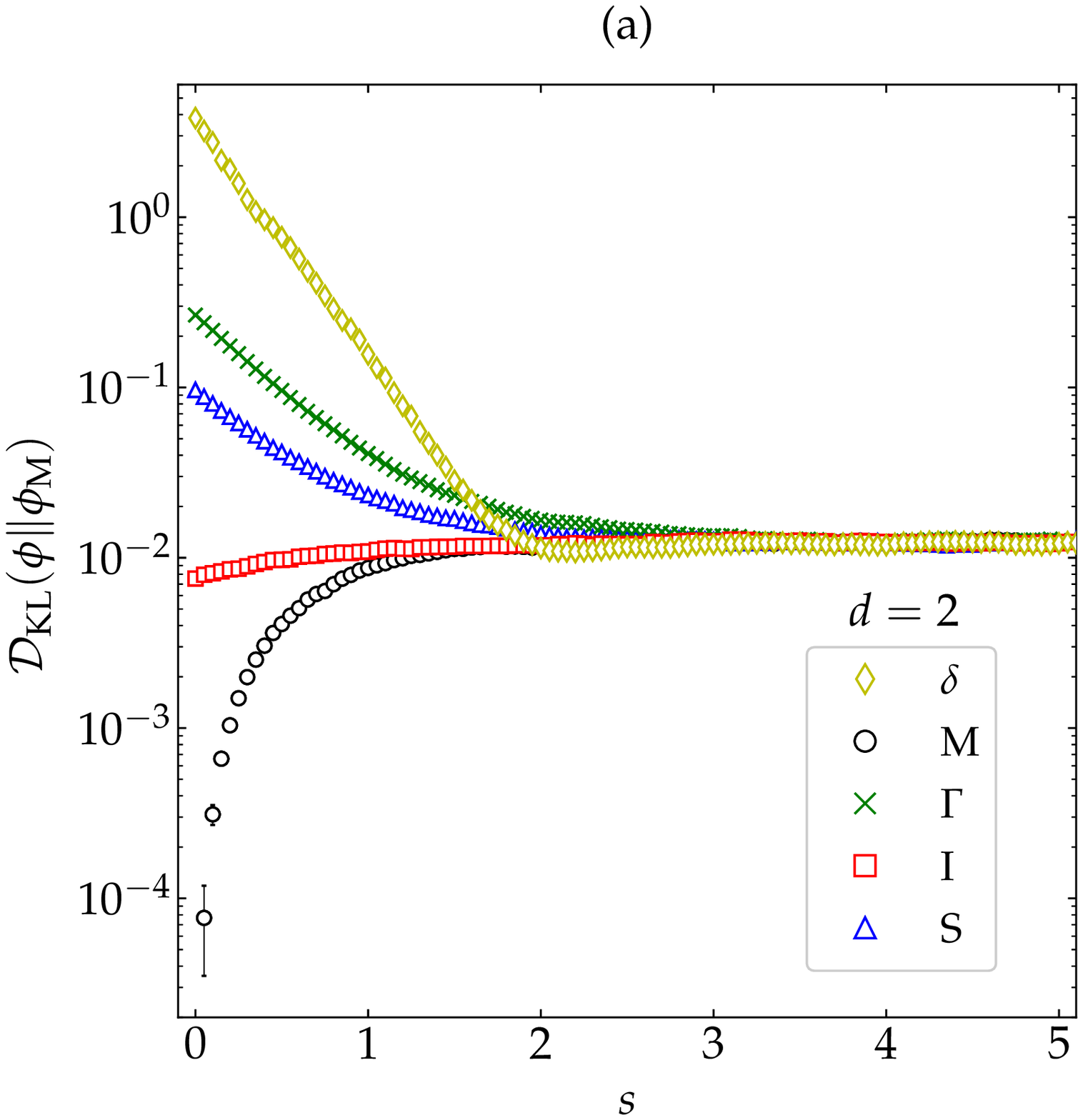}\hspace{.9cm}\includegraphics[width=.45\columnwidth]{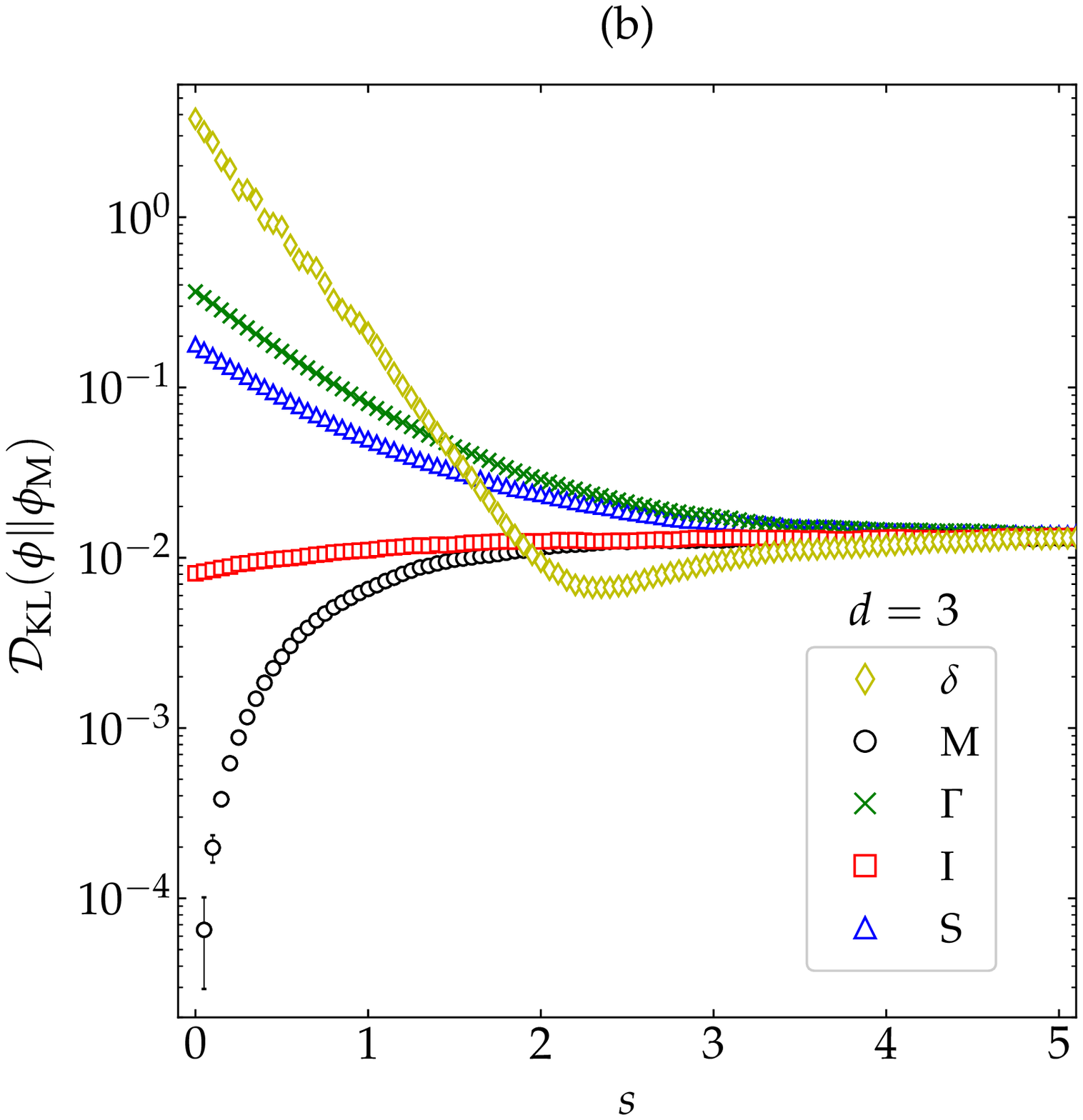}
\caption{Evolution of $\mathcal{D}_{\text{KL}}(\phi\|\phi_\text{M})$
(in logarithmic scale) for a coefficient of restitution  $\alpha=0.1$
as a function of the average number of collisions per particle for hard (\textbf{a}) disks and (\textbf{b}) spheres. Symbols represent MD simulation results.
Five different initial conditions are considered (see Appendix \ref{app:C}): $\delta$~($\diamond$), M ($\circ$), $\Gamma$ ($\times$), I ($\square$), and~S ($\triangle$). The~error bars are smaller than the size of the symbols, except~when $\mathcal{D}_{\text{KL}}(\phi\|\phi_{\text{M}})\lesssim 10^{-4}$ for the initial condition M.\label{fig:DKL_ev_M_conds}}
\end{figure}

In order to examine how generic the non-monotonic behavior of $\mathcal{D}_{\text{KL}}(\phi\|\phi_{\text{M}})$ is for high inelasticity, we have taken the case $\alpha=0.1$ and considered the same five different initial conditions as in \mbox{Figures~\ref{fig:a2Evol_new} and \ref{fig:a3Evol_new}} (\mbox{see Appendix \ref{app:C}}).
The results are displayed in Figure~\ref{fig:DKL_ev_M_conds}, where we can observe that only the initial condition $\delta$ exhibits a non-monotonic behavior, whereas $\mathcal{D}_{\text{KL}}(\phi\|\phi_{\text{M}})$ decays (grows) monotonically in the cases of the initial conditions $\Gamma$ and S (M and I). This shows that the nonmoniticity in the time evolution of $\mathcal{D}_{\text{KL}}(\phi\|\phi_{\text{M}})$ is a rather subtle effect requiring high inelasticity and special initial~conditions.
\unskip

\subsubsection{HCS Distribution as a Reference ($\phi_\refe=\phi_{\hcs}$)}

By using formal arguments from Refs.\ \cite{MMR06,MM06,MM09}, Garc\'ia de Soria~et~al.\ \cite{GMMMRT15} proved by means of a perturbation analysis around $\alpha=1$ that $\phi_{\hcs}$ is a unique local minimizer of the entropy production, implying that $\partial_s  \mathcal{D}_{\text{KL}}(\phi\|\phi_{\hcs})\leq 0$, in~the \emph{quasielastic} limit. Those authors also conjectured that this result keeps being valid in the whole inelasticity regime, this conjecture being  supported by simulations for $\alpha\geq 0.8$ in the freely cooling~case.

\begin{figure}[H]
\centering
\includegraphics[width=.45\columnwidth]{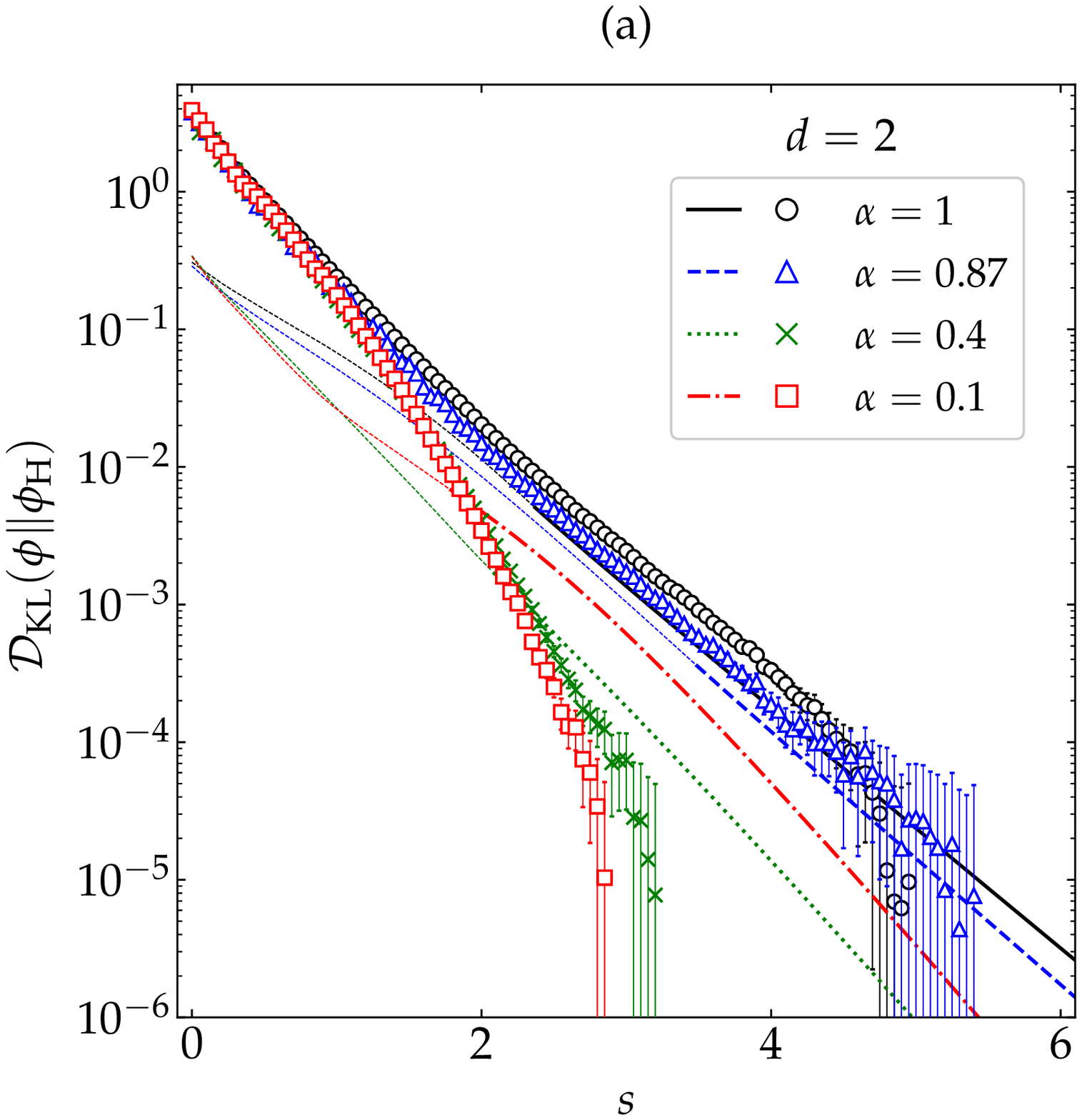}\hspace{1cm}\includegraphics[width=.45\columnwidth]{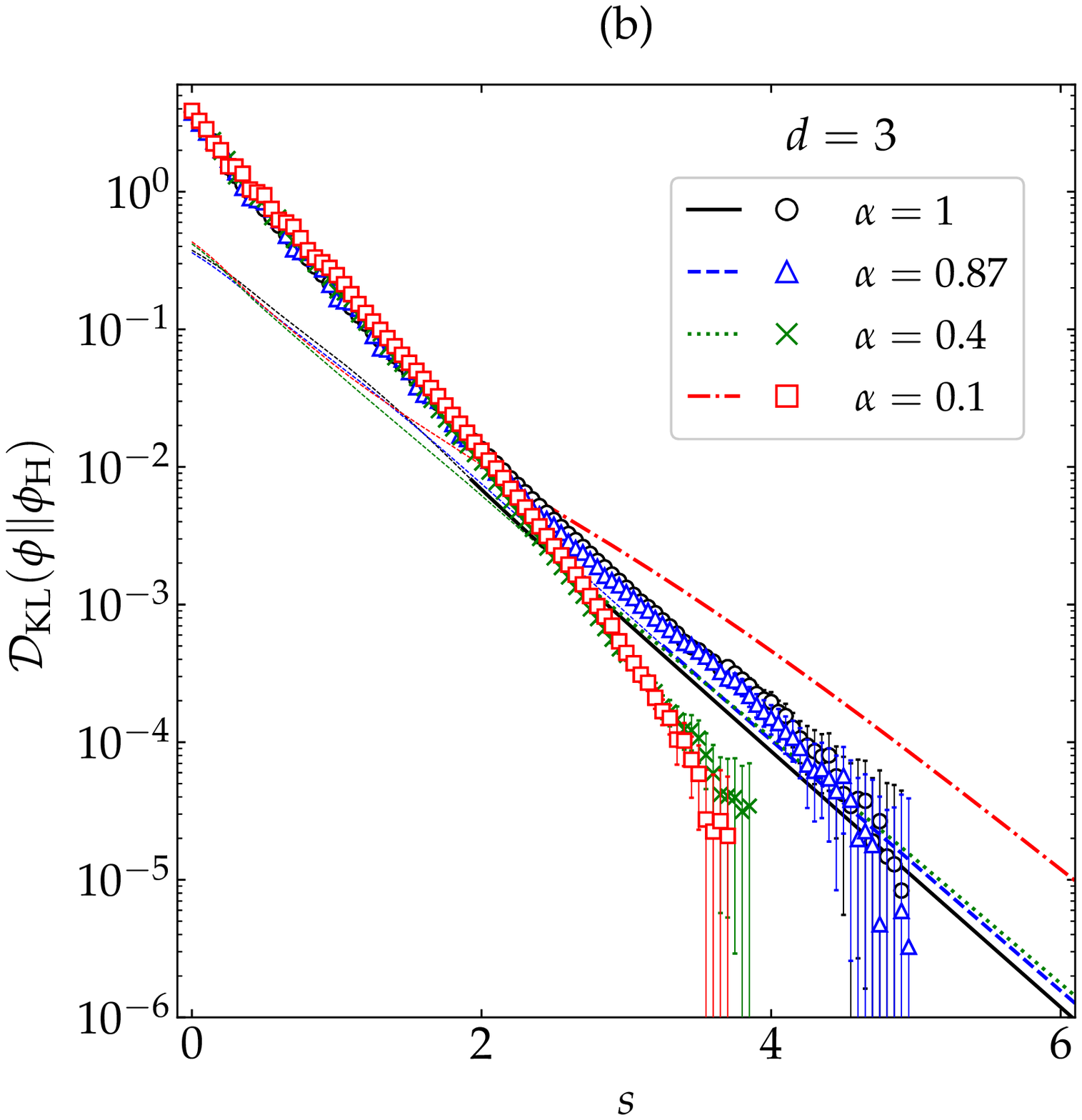}
\caption{Evolution of $\mathcal{D}_{\text{KL}}(\phi\|\phi_\hcs)$
(in logarithmic scale)
as a function of the average number of collisions per particle for (\textbf{a}) disks and (\textbf{b}) spheres. Symbols represent MD simulation results, while the lines correspond to the theoretical prediction \eqref{DKLH} (the thin dashed lines for the first stage of the evolution meaning that it was necessary to take the real part). The~values of the  coefficient of restitution are  $\alpha=0.1$ ($\square$), $0.4$ ($\times$), $0.87$ ($\triangle$), and~$1$ ($\circ$) The error bars in the simulation data are smaller than the size of the symbols, except~when $\mathcal{D}_{\text{KL}}(\phi\|\phi_{\text{M}})\lesssim 10^{-4}$.\label{fig:DKL_ev_HCS}}
\end{figure}

By performing MD simulations for a wide range of inelasticities ($\alpha = 0.1$, $0.2$, $0.3$, $0.4$, $0.5$, $0.6$, $1/\sqrt{2}$, $0.8$, $0.87$, $0.95$, and~$0.99$), we have found further support for the  inequality  $\partial_s  \mathcal{D}_{\text{KL}}(\phi\|\phi_{\hcs})\leq 0$. As~an illustration, Figure~\ref{fig:DKL_ev_HCS} shows the evolution of $\mathcal{D}_{\text{KL}}(\phi\|\phi_{\hcs})$ for $\alpha = 0.1$, $0.4$, $0.87$, and~$1$, starting from the same initial states as in Figures~\ref{fig:a2Evol}, \ref{fig:a3Evol}, and~\ref{fig:DKL_M_evl}.
In the evaluation of $\mathcal{D}_{\text{KL}}(\phi\|\phi_{\hcs})$, we have used the simulation results for both the transient distribution $\phi(\cc;s)$ and the asymptotic HCS distribution $\phi_\hcs(\cc)$ (see Appendix \ref{app:A}).
Our MD results are compared with a theoretical approximation similar to that of Equation \eqref{DKLM}, i.e.,
\begin{equation}
\label{DKLH}
\mathcal{D}_{\text{KL}}(\phi\|\phi_{\hcs})\approx\int\dif\cc\,\phi_{\text{M}}(\cc)\left[1+ a_2(s)S_2(c^2)+ a_3(s)S_3(c^2)\right]\ln \frac{1+ a_2(s)S_2(c^2)+ a_3(s)S_3(c^2)}{1+ a_2^\hcs S_2(c^2)+ a_3^\hcs S_3(c^2)},
\end{equation}
where  again the real part of the right-hand side is taken if $1+ a_2(s)S_2(c^2)+ a_3(s)S_3(c^2)<0$ for a certain range of velocities.
The results (both from MD and from the approximate theory) displayed in Figure~\ref{fig:DKL_ev_HCS}  show that $\mathcal{D}_{\text{KL}}(\phi\|\phi_{\hcs})$ indeed decays monotonically to $0$, even for very strong inelasticity, thus supporting its status as  a very sound candidate of Lyapunov functional. It is also interesting to note that the characteristic relaxation time is generally shorter for disks than for spheres and tends to decrease with increasing~inelasticity.

In order to reinforce the monotonic decay of $\mathcal{D}_{\text{KL}}(\phi\|\phi_{\hcs})$ observed in Figure~\ref{fig:DKL_ev_HCS} for several representative values of the coefficient of restitution, let us now take the most demanding case ($\alpha=0.1$) and choose the five initial conditions already considered in Figures~\ref{fig:a2Evol_new}, \ref{fig:a3Evol_new}, and~\ref{fig:DKL_ev_M_conds} (see Appendix~\ref{app:C}).
Figure~\ref{fig:DKL_ev_HCS_conds} shows that the evolution of $\mathcal{D}_{\text{KL}}(\phi\|\phi_{\hcs})$ keeps being monotonic for this wide spectrum of representative initial conditions, the~relaxation to the HCS being again faster for disks than for spheres.
It is also interesting to comment that, although~the largest initial divergence corresponds to the initial distribution  $\delta$, this divergence decays more rapidly than the other four ones, and~even seems to overtake the divergence associated with the initial condition $\Gamma$.

\begin{figure}[H]
\centering
\includegraphics[width=.45\columnwidth]{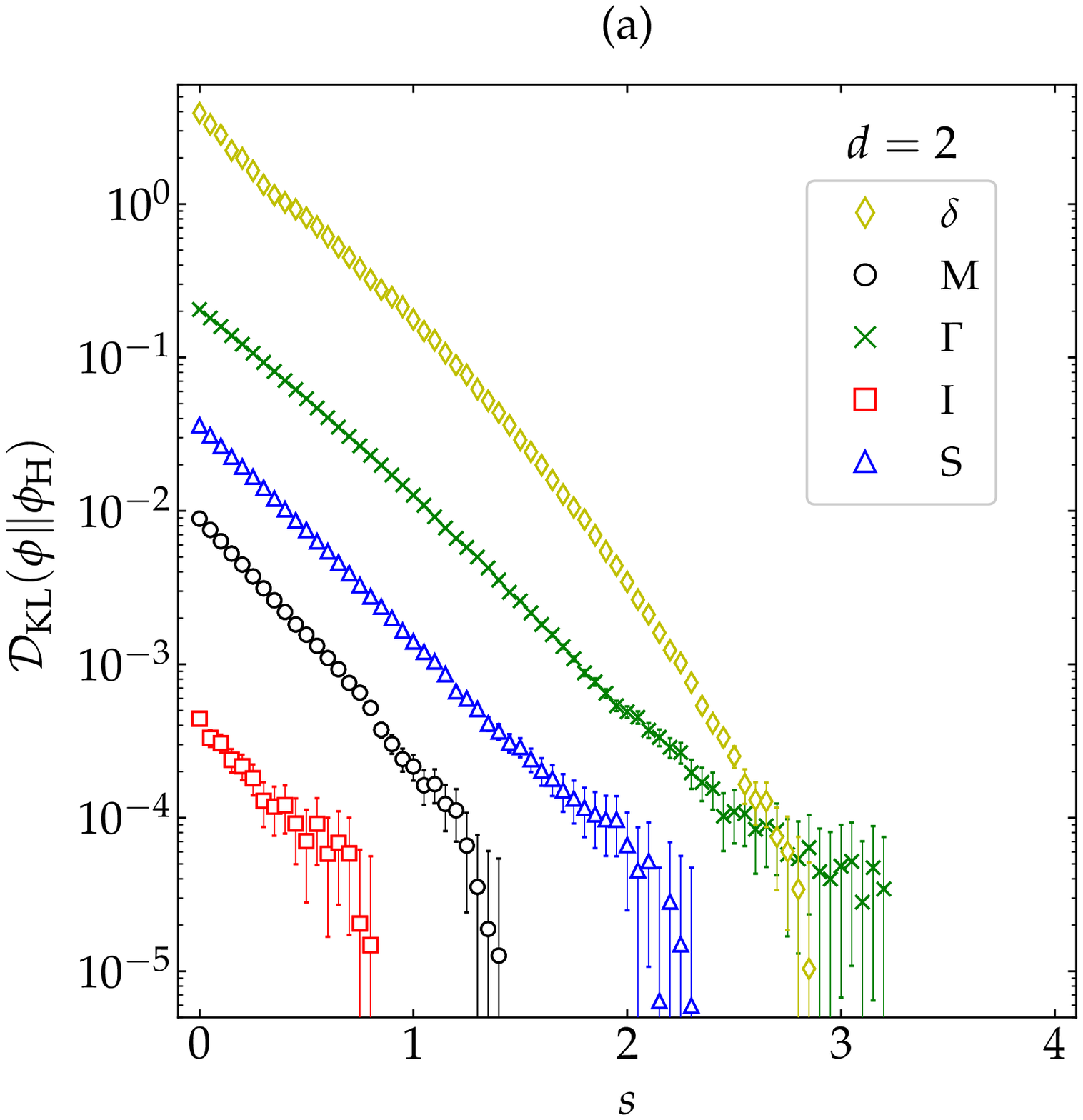}\hspace{1cm}\includegraphics[width=.45\columnwidth]{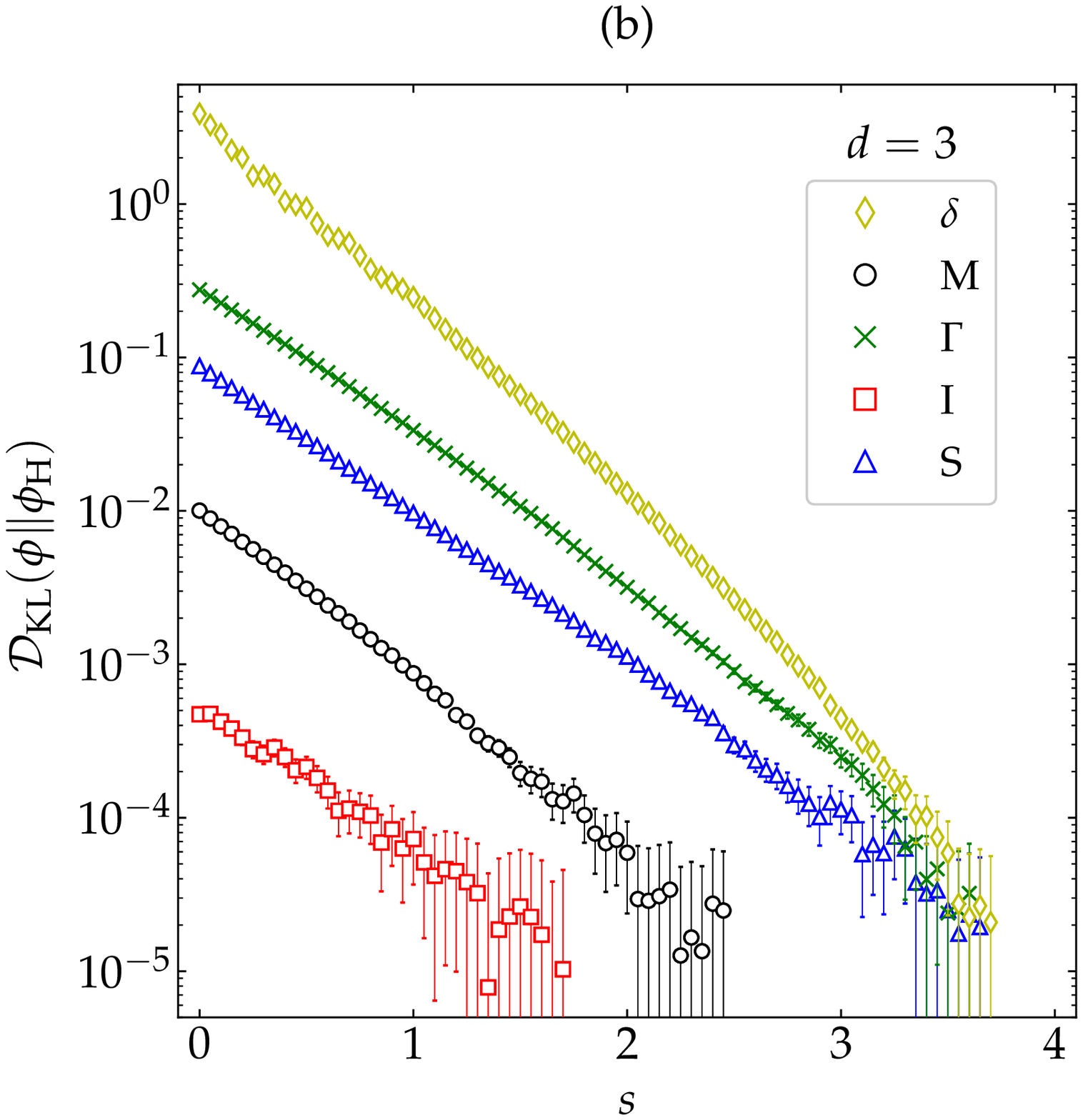}
\caption{Evolution of $\mathcal{D}_{\text{KL}}(\phi\|\phi_\hcs)$
(in logarithmic scale) for a coefficient of restitution  $\alpha=0.1$
as a function of the average number of collisions per particle for hard (\textbf{a}) disks and (\textbf{b}) spheres. Symbols represent MD simulation results.
Five different initial conditions are considered (see Appendix \ref{app:C}): $\delta$~($\diamond$), M ($\circ$), $\Gamma$ ($\times$), I ($\square$), and~S ($\triangle$). The~error bars are smaller than the size of the symbols, except~when $\mathcal{D}_{\text{KL}}(\phi\|\phi_{\text{M}})\lesssim 10^{-4}$. \label{fig:DKL_ev_HCS_conds}}
\end{figure}

While a rigorous mathematical proof of $\partial_s  \mathcal{D}_{\text{KL}}(\phi\|\phi_{\hcs})\leq 0$ is still lacking (see, however, Ref.~\cite{P04} for the sketch of a proof in the context of the linear Boltzmann equation), we will now prove this inequality by using a simplified \emph{toy model}.
We start from the infinite series expansion \eqref{eq:fv} and imagine a formal bookkeeping parameter $\epsilon$ in front of the Sonine summation. Then, to~the second order in $\epsilon$,

\small\bal
\frac{\phi(\cc;s)}{\phi_{\text{M}}(\cc)}\ln\frac{\phi(\cc;s)}{\phi_\hcs(\cc)}=&\epsilon \sum_{k=2}^\infty\left[a_k(s)-a_k^\hcs\right]S_k(c^2)
+\frac{\epsilon^2}{2}\sum_{k,k'=2}^\infty \left[a_k(s)-a_k^\hcs\right]\left[a_{k'}(s)-a_{k'}^\hcs\right]S_k(c^2)S_{k'}(c^2)\nn
&+\mathcal{O}(\epsilon^3).
\eal
Next, taking into account the orthogonality condition \eqref{ortho}, we get
\begin{subequations}
\begin{equation}
\mathcal{D}_{\text{KL}}(\phi\|\phi_{\hcs})=\frac{\epsilon^2}{2}\sum_{k}^\infty \mathcal{N}_k\left[a_k(s)-a_k^\hcs\right]^2+\mathcal{O}(\epsilon^3),
\end{equation}
\begin{equation}
\partial_s\mathcal{D}_{\text{KL}}(\phi\|\phi_{\hcs})={\epsilon^2}\sum_{k}^\infty \mathcal{N}_k\left[a_k(s)-a_k^\hcs\right]\partial_s a_k(s)+\mathcal{O}(\epsilon^3).
\end{equation}
\end{subequations}
Interestingly, this approximation preserves the positive-definiteness of the KLD. Note also that, to~order $\epsilon^2$, $\mathcal{D}_{\text{KL}}(\phi\|\phi_{\hcs})$ is symmetric under the exchange $\phi\leftrightarrow\phi_{\hcs}$, i.e.,~$\mathcal{D}_{\text{KL}}(\phi\|\phi_{\hcs})-\mathcal{D}_{\text{KL}}(\phi_\hcs\|\phi)=\mathcal{O}(\epsilon^3)$. Finally, consistent with the derivation of Equations \eqref{a2b} and \eqref{dsa2}, we neglect the cumulants $a_k$ with $k\geq 3$ and apply Equation \eqref{dsa2} to obtain
\begin{subequations}
\label{toy}
\begin{equation}
\label{toy_a}
\mathcal{D}_{\text{KL}}(\phi\|\phi_{\hcs})\approx \frac{d(d+2)}{16}\left[a_2(s)-a_2^\hcs\right]^2,
\end{equation}
\begin{equation}
\label{toy_b}
\partial_s\mathcal{D}_{\text{KL}}(\phi\|\phi_{\hcs})\approx -\frac{d(d+2)}{8}K_2\left[1+a_2(s)\right]\left[a_2(s)-a_2^\hcs\right]^2\leq 0,
\end{equation}
\end{subequations}
where we have formally set $\epsilon=1$. Although~a certain number of approximations have been done to derive the toy model \eqref{toy}, it undoubtedly provides further support to the conjecture $\partial_s\mathcal{D}_{\text{KL}}(\phi\|\phi_{\hcs})\leq 0$.

\subsubsection{Relative Entropy of $\phi_\hcs$ with Respect to $\phi_{\text{M}}$}

It is well known that, in~a freely cooling granular gas, the~HCS VDF   is generally close (at least within the range of thermal velocities) to a Maxwellian. In~particular, the~cumulants $a_k^\hcs$ are rather small in magnitude, except~at large inelasticity (see Figure~\ref{fig:a2a3st}). On~the other hand, the~HCS VDF exhibits an exponential high-velocity tail, $\ln \phi_\hcs(\cc)$$\sim$$-c$, with~respect to the Maxwellian behavior, $\ln \phi_{\text{M}}(\cc)$$\sim$$-c^2$ \cite{YSS20,EP97,vNE98}.

Here, we have one more tool to measure how far $\phi_{\text{M}}(\cc)$ is from $\phi_\hcs(\cc)$, namely the  KLD  from $\phi_{\text{M}}$ to $\phi_\hcs$ (or~relative entropy of $\phi_\hcs$ with respect to $\phi_{\text{M}}$), i.e.,~$ \mathcal{D}_{\text{KL}}(\phi_\hcs\|\phi_{\text{M}})$. Note, however, that, as~said at the beginning of this section, the~KLD is not a real metric since it does not fulfill either symmetry or triangle inequality properties of a~distance.

\begin{figure}[H]
\centering
\includegraphics[width=.45\columnwidth]{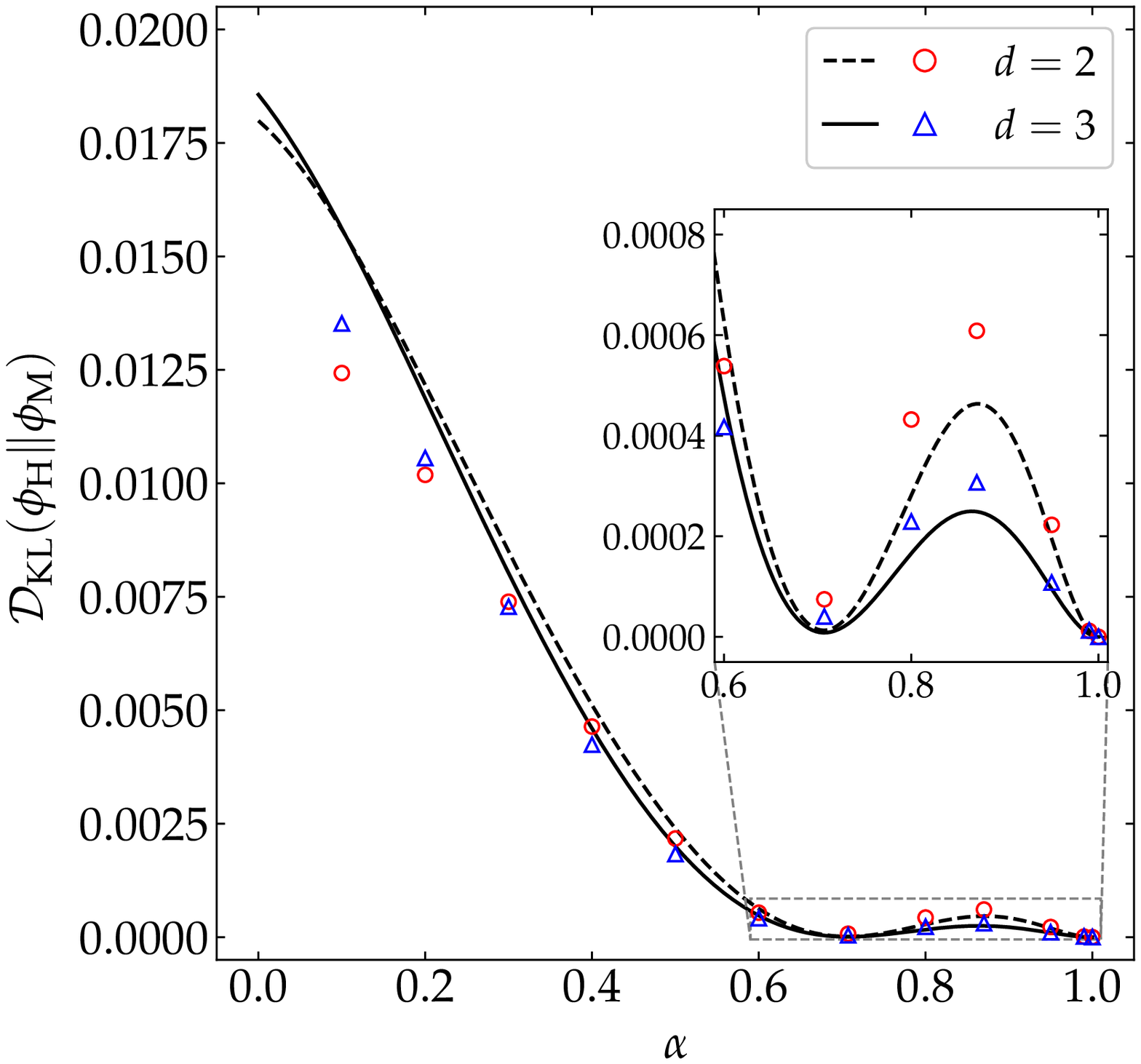}
\caption{Plot of $\mathcal{D}_{\text{KL}}(\phi_{\hcs}\|\phi_{M})$ as a function of the coefficient of restitution $\alpha$ for disks (-- --, $\circ$) and spheres (---, $\triangle$). Symbols represent MD simulation results, while the lines correspond to the theoretical prediction provided by Equation \eqref{DKLM} with $a_2(s)\to a_2^\hcs$ and $a_3(s)\to a_3^\hcs$. The~inset magnifies the
 region $0.6\leq\alpha\leq 1$. The~error bars in the simulation data are smaller than the size of the symbols.\label{fig:DKL_st}}
\end{figure}

Figure~\ref{fig:DKL_st} displays the $\alpha$-dependence of $\mathcal{D}_{\text{KL}}(\phi_{\hcs}\|\phi_{\text{M}})$ for both disks and spheres, as~obtained from our MD simulations (see again Appendix \ref{app:A}) and from the simple estimate \eqref{DKLM} with $a_2(s)\to a_2^\hcs$ and $a_3(s)\to a_3^\hcs$. We~can observe that the theoretical truncated approach successfully captures (i) a weak influence of dimensionality (in contrast to the fourth and sixth cumulants  plotted in Figure~\ref{fig:a2a3st}), (ii) a crossover from
\mbox{$\left.\mathcal{D}_{\text{KL}}(\phi_{\hcs}\|\phi_{\text{M}})\right|_{d=2}<\left.\mathcal{D}_{\text{KL}}(\phi_{\hcs}\|\phi_{\text{M}})\right|_{d=3}$} for very large inelasticity to $\left.\mathcal{D}_{\text{KL}}(\phi_{\hcs}\|\phi_{\text{M}})\right|_{d=2}>\left.\mathcal{D}_{\text{KL}}(\phi_{\hcs}\|\phi_{\text{M}})\right|_{d=3}$ for smaller inelasticity, and~(ii) a non-monotonic dependence on $\alpha$, with~a (small but nonzero) local minimum at about $\alpha=1/\sqrt{2}\simeq 0.71$ and a local maximum at about $\alpha=0.87$. The~latter property implies that, in~the region $0.6\lesssim \alpha<1$, three systems differing in the value of $\alpha$ may share the same divergence of $\phi_{\text{M}}$ from $\phi_\hcs$. The~qualitative shape of $\mathcal{D}_{\text{KL}}(\phi_{\hcs}\|\phi_{\text{M}})$ as a function of $\alpha$ agrees with a toy model analogous to that of Equation~\eqref{toy_a}, namely $\mathcal{D}_{\text{KL}}(\phi_{\hcs}\|\phi_{\text{M}})\approx \frac{d(d+2)}{16}{a_2^\hcs}^2$.

\vskip 0.3cm

%%%%%%%%%%%%%%%%%%%%%%%%%%%%%%%%%%%%%%%%%%
\section{Summary and~Conclusions}
\label{sec5}

In this work, we have mainly focused on the role as a potential entropy-like Lyapunov  functional played by the KLD of a reference VDF ($\phi_\refe$) with respect to the \emph{spatially homogeneous} time-dependent VDF ($\phi)$, i.e.,~ $\DKL(\phi\|\phi_\refe)$, as~supported by MD simulations in a freely cooling granular-gas~model.

First, we have revisited the problem of obtaining, by~kinetic theory methods, simple approximations for the HCS fourth ($a_2^\hcs$) and sixth ($a_3^\hcs$) cumulants, and~have derived explicit time-dependent solutions, $a_2(s)$ and $a_3(s)$, for~arbitrary (homogeneous) initial conditions. Comparison with our MD results shows an excellent general performance of $a_2^\hcs$ and $a_2(s)$ for values of the coefficient of restitution as low as $\alpha=0.1$ and for a variety of initial conditions. In~the case of the sixth cumulant, however, the~agreement is mainly semi-quantitative. In~any case, our~MD data for $a_2^\hcs$ and $a_3^\hcs$ agree very well with previous simulations of the inelastic Boltzmann equation~\cite{BRC96,MS00,BP06,SM09}, thus validating the applicability of kinetic theory (including the Stosszahlansatz) even for high inelasticity.
We emphasize that, to~the best of our knowledge, such a comprehensive MD analysis of the fourth and sixth cumulants had not been carried out before. We are not aware either of a previous (approximate) theoretical derivation of the time-dependent quantities $a_2(s)$ and $a_3(s)$.

As a first candidate to a Lyapunov functional, we have considered the KLD with a Maxwellian reference VDF ($\phi_\refe=\phi_{\text{M}}$). However,~this possibility is clearly discarded as both simulation and a simple theoretical approach show that $\DKL(\phi\|\phi_\text{M})$ does not relax monotonically for highly inelastic systems and certain initial conditions. On~the other hand, when the asymptotic HCS VDF is chosen as a reference ($\phi_\refe=\phi_{\hcs}$), the~results show that the relaxation of $\DKL(\phi\|\phi_\hcs)$ is monotonic for a wide spectrum of inelasticities and initial conditions. This is further supported by a simplified toy model, according to which $\partial_s \DKL(\phi\|\phi_\hcs)$$\sim$$-[a_2(s)-a_2^\hcs]^2\leq 0$.
While simulation results supporting the conjecture  $\partial_s \DKL(\phi\|\phi_\hcs)\leq 0$ had been presented before~\cite{GMMMRT15}, it is subjected here to more stringent tests by considering highly dissipative collisions ($\alpha=0.1$ and $0.4$) and a repertoire of different initial conditions. In~fact, it is only under those more extreme conditions when one can reject the Maxwellian  as a proper candidate for the reference~VDF.

We have also used $\mathcal{D}_{\text{KL}}(\phi_{\hcs}\|\phi_{\text{M}})$ to characterize the departure of the Maxwellian distribution as an approximation to the actual HCS distribution. Interestingly, we found a non-monotonic influence of the coefficient of restitution on $\mathcal{D}_{\text{KL}}(\phi_{\hcs}\|\phi_{\text{M}})$, with~a (nonzero) local minimum at $\alpha\simeq 1/\sqrt{2}\simeq 0.71$ and a (small) local maximum at $\alpha\simeq 0.87$. This non-monotonicity implies a \emph{degeneracy} of $\mathcal{D}_{\text{KL}}(\phi_{\hcs}\|\phi_{\text{M}})$ in the sense that three different coefficients of restitution (within the region $0.6\lesssim\alpha <1$) may share a common value of the KLD from $\phi_{\text{M}}$ to $\phi_{\hcs}$.
The analysis of $\mathcal{D}_{\text{KL}}(\phi_{\hcs}\|\phi_{\text{M}})$ is an additional asset of our~work.

We expect that the results presented in this paper may stimulate further studies on the quest of proving (or disproving, if~a counterexample is found) the extension of Boltzmann's celebrated $H$-theorem to the realm of dissipative inelastic collisions in homogeneous states.
In this respect, it must be remarked that, since the simulation results we have presented are obtained from the MD technique (which numerically solves Newton's equations of motion) and not from the DSMC method (which numerically solves the Boltzmann equation), it~is not obvious from a strict mathematical point of view that the obtained results imply the decay of the KLD in the context of the Boltzmann equation. On~the other hand, on~physical grounds, it is expected that such an implication~holds.

As a final remark, it is worth emphasizing that, even if some kind of generalized $H$-theorem could be proved for homogeneous states, its extension to inhomogeneous situations would be far from trivial since the HCS is unstable under long-wavelength~perturbations.

%%%%%%%%%%%%%%%%%%%%%%%%%%%%%%%%%%%%%%%%%%
\vspace{6pt}

\authorcontributions{A.S. proposed the idea and A.M. carried out the simulations.
Both authors participated in the analysis and discussion of the results and worked on the revision and writing of the final~manuscript. All~authors have read and agreed to the published version of the~manuscript.}

%%%%%%%%%%%%%%%%%%%%%%%%%%%%%%%%%%%%%%%%%%
\funding{The authors acknowledge financial support from the Spanish Agencia Estatal de Investigaci\'on through Grant No. FIS2016-76359-P and the Junta de Extremadura
(Spain) through Grant No. GR18079, both partially financed by Fondo Europeo de Desarrollo Regional funds.
A.M. is grateful to the Spanish Ministerio de Ciencia, Innovaci\'on y Universidades for a predoctoral fellowship~FPU2018-3503.}

\conflictsofinterest{The authors declare no conflict of interest. %The funding sponsors had no role in the design of the study; in the collection, analyses, or~interpretation of data; in the writing of the manuscript, and~in the decision to publish the results''.
}

\abbreviations{The following abbreviations are used in this manuscript:\\

\noindent
\begin{tabular}{@{}ll}
DSMC  & Direct simulation Monte Carlo\\
HCS&Homogenous cooling state\\
KLD & Kullback--Leibler divergence\\
MD  & Molecular dynamics\\
VDF&Velocity distribution function\\
\end{tabular}}

%%%%%%%%%%%%%%%%%%%%%%%%%%%%%%%%%%%%%%%%%%
%% optional
%\supplementary{The following are available online at \linksupplementary{s1}, Figure S1: title, Table S1: title, Video S1: title.}

% Only for the journal Methods and Protocols:
% If you wish to submit a video article, please do so with any other supplementary material.
% \supplementary{The following are available at \linksupplementary{s1}, Figure S1: title, Table S1: title, Video S1: title. A supporting video article is available at doi: link.}

%%%%%%%%%%%%%%%%%%%%%%%%%%%%%%%%%%%%%%%%%%
%% optional
\appendixtitles{yes} %Leave argument "no" if all appendix headings stay EMPTY (then no dot is printed after "Appendix A"). If~the appendix sections contain a heading then change the argument to "yes".
\appendix
\section{Simulation and Numerical~Details}
\label{app:A}
\unskip
Event-driven MD simulations  were carried out using the DynamO software~\cite{BSL11} on $N$ particles in a $d$-dimensional cubic box of side $L$ with periodic boundary conditions. We chose $(N,L/\sigma)=(10^4,4\,641.58)$ and $(1.35\times 10^4,407.16)$ for disks and spheres, respectively. Thus, the~associated number densities were $n\sigma^2=4.64\times 10^{-4}$ (disks) and $n\sigma^3=2.00\times 10^{-4}$ (spheres).
The critical lengths for the development of instabilities at those densities~\cite{G19} are estimated to be $L_c/\sigma\approx 8.3\times 10^3$ (disks) and $ 1.3\times 10^4$ (spheres) for the most demanding case ($\alpha=0.1$). This represents ratios $L_c/L\approx 1.8$ (disks) $L_c/L\approx 32.8$ (spheres). Those ratios generally increase with decreasing inelasticity. For~instance, at~$\alpha=0.87$, one finds $L_c/L\approx 3.4$ (disks) and $ 62.5$ (spheres). Therefore, the~simulations are performed in the region of parameters where the systems are~stable.

Since the DynamO code is designed for three-dimensional setups, we used it for the two-dimensional case by imposing a coordinate $z=0$ to every particle and carefully avoiding any overlap in the initial ordered arrangement. The~system melted very quickly and no inhomogeneities were observed thereafter. A~velocity rescaling was done periodically in order to avoid numerical errors due to the cooling process and extremely small~numbers.

To represent the VDF and the KLD in simulations, let us first introduce the probability distribution function of the velocity modulus,
\begin{equation}
\label{A_1}
\Phi(c;s)=c^{d-1}\int\dif \widehat{\cc}\,\phi(\cc;s)=\Omega_d c^{d-1}\phi(\cc;s),\quad \Omega_d\equiv \frac{2\pi^{\dt/2}}{\Gamma(\dt/2)},
\end{equation}
where in the second step we have assumed that the VDF $\phi(\cc;s)$ is isotropic and $\Omega_d$ is the $d$-dimensional solid angle.
Thus, Equation \eqref{KLD} can be rewritten as
\begin{equation}
\label{KLD_Phi}
\mathcal{D}_{\text{KL}}(\phi\|\phi_\refe)=\int_0^\infty \dif c\medspace  \Phi(c;s)\ln\frac{\Phi(c;s)}{\Phi_\refe(c)}.
\end{equation}

The functions $\Phi(c;s)$ and $\Phi_\hcs(c)$ are numerically approximated by a discrete histogram, with~a certain constant bin width     $\hh$, i.e.,
\begin{equation}
\label{eq:disc_hist}
    \Phi(c_i;s)\approx\frac{N_i(s)}{N\hh}, \quad \Phi_\hcs(c_i)\approx\frac{N_i^\hcs}{N\hh},\quad c_i=\left(i-\frac{1}{2}\right)\hh,\quad i=1,2,\ldots,M.
\end{equation}
Here, $N_i(s)$ is the number of particles with a speed $c$ inside the interval  $c_i-\hh/2\leq c<c_i+\hh/2$, $N_i^\hcs$ is evaluated by averaging $N_i(s)$ between $s=10$ to $s=40$ with a timestep $\delta s=0.2$, and~$M$ is the total number of bins considered.
In consistency with Equation \eqref{eq:disc_hist}, the~Maxwellian VDF is also discretized as
\bal
\label{eq:max_appr}
\Phi_{\text{M}}(c_i)\approx &\frac{\pi^{-d/2}\Omega_d}{\hh}\int_{c_i-\hh/2}^{c_i+\hh/2} \dif c\, c^{d-1}e^{-c^2}\nn
=&\begin{cases}
  \frac{e^{-\left(c_i-\frac{\hh}{2}\right)^2}-e^{-\left(c_i+\frac{\hh}{2}\right)^2}}{\hh},\quad (d=2),\\
  \frac{\mathrm{erf}\left(c_i+\frac{\hh}{2}\right)-\mathrm{erf}\left(c_i-\frac{\hh}{2}\right)}{\hh}+\frac{2}{\sqrt{\pi}}\frac{
 \left(c_i-\frac{\hh}{2}\right) e^{-\left(c_i-\frac{\hh}{2}\right)^2}-\left(c_i+\frac{\hh}{2}\right)e^{-\left(c_i+\frac{\hh}{2}\right)^2}}{\hh},\quad (d=3),
\end{cases}
\eal
where $\mathrm{erf}(x)=\frac{2}{\sqrt{\pi}}\int_0^x \dif t\medspace e^{-t^2}$ is the error~function.

Next, the~KLD \eqref{KLD_Phi} with $\phi_\refe(\cc)=\phi_{\text{M}}(\cc)$ and with $\phi_\refe(\cc)=\phi_{\hcs}(\cc)$ are approximated in the simulations by
\begin{equation}
   \mathcal{D}_{\text{KL}}(\phi\|\phi_{\text{M}})\approx \sum_{i=1}^M  \frac{N_i(s)}{N}\ln \frac{N_i(s)/N\hh}{\Phi_{\text{M}}(c_i)} ,\quad
     \mathcal{D}_{\text{KL}}(\phi\|\phi_{\hcs})\approx \sum_{i=1}^M  \frac{N_i(s)}{N}\ln \frac{N_i(s)}{N_i^\hcs},
\end{equation}
where $\Phi_{\text{M}}(c_i)$ is given by Equation  \eqref{eq:max_appr}.
Analogously,
\begin{equation}
        \mathcal{D}_{\text{KL}}(\phi_\hcs\|\phi_{\text{M}})\approx \sum_{i=1}^M  \frac{N_i^\hcs}{N}\ln \frac{N_i^\hcs/N\hh}{\Phi_{\text{M}}(c_i)}.
\end{equation}

A comment is now in order. In~the case of elastic collisions ($\alpha=1$), one obviously should have \mbox{$\Phi_\hcs (c_i)=\Phi_{\text{M}}(c_i)$} and hence $\left.\mathcal{D}_{\text{KL}}(\phi_\hcs\|\phi_{\text{M}})\right|_{\alpha=1}=0$. However, since $\Phi_\hcs (c_i)$ is evaluated in simulations by Equation \eqref{eq:disc_hist} for any $\alpha$, the~equality $\Phi_\hcs (c_i)=\Phi_{\text{M}}(c_i)$ for $\alpha=1$ is not identically verified bin to bin due to fluctuations. As~a consequence, in~the simulations, $\left.\mathcal{D}_{\text{KL}}(\phi_\hcs\|\phi_{\text{M}})\right|_{\alpha=1}\sim 10^{-5}\neq 0$. This is an unavoidable background noise that was subtracted from the KLD obtained by simulations, i.e.,~$\mathcal{D}_{\text{KL}}(\phi\|\phi_\refe)\to \mathcal{D}_{\text{KL}}(\phi\|\phi_\refe)-\left.\mathcal{D}_{\text{KL}}(\phi_\hcs\|\phi_{\text{M}})\right|_{\alpha=1}$.

We have chosen the values $\hh=0.03$ and $M=200$. The~results presented in the main text for any given quantity are obtained by averaging over $50$ independent~realizations.

\section{Initial~Conditions}
\label{app:C}
For the analysis of the evolution of $a_2(s)$, $a_3(s)$, $\DKL(\phi\|\phi_\text{M})$, and~$\DKL(\phi\|\phi_\hcs)$ with $\alpha=0.1$, we have chosen five different initial conditions. The~first one is the same as considered in Figures~\ref{fig:a2Evol}, \ref{fig:a3Evol}, \ref{fig:DKL_M_evl}, and~\ref{fig:DKL_ev_HCS}, i.e.,~an ordered crystalized configuration with isotropic velocities of a common magnitude. In~terms of the distribution defined by Equation \eqref{A_1}, this initial condition reads
\begin{equation}
\label{C_1}
\Phi_\delta(c)=\delta\left(c-\sqrt{{d}/{2}}\right),
\end{equation}
which will be labeled with the Greek letter $\delta$. The~second initial distribution is just a Maxwellian (label M), i.e.,
\begin{equation}
\Phi_\text{M}(c)=\frac{2}{\Gamma(\frac{d}{2})}c^{d-1}e^{-c^2}.
\end{equation}
Next, we choose the gamma distribution (label $\Gamma$) normalized to $\braket{c^2}=\frac{d}{2}$, namely
\begin{equation}
\Phi_\Gamma(c)=\frac{2}{\theta^{\frac{d}{2\theta}}\Gamma(\frac{d}{2\theta})}c^{d/\theta-1}e^{-c^2/\theta},
\end{equation}
where $\theta>0$ can be freely chosen. The~fourth- and sixth-order moments are $\braket{c^4}=\frac{d(d+2\theta)}{4}$ and \mbox{$\braket{c^6}=\frac{d(d+2\theta)(d+4\theta)}{8}$}, so that $a_2=\frac{2(\theta-1)}{d+2}$ and $a_3=-\frac{8(\theta-1)(\theta-2)}{(d+2)(d+4)}$. Here, we have taken $\theta=2.16$ and $2.45$ for $d=2$ and $3$, respectively.

The remaining two initial conditions are prepared by applying a coefficient of normal restitution $\alpha_0$ and allowing the system to reach the corresponding steady state (in the scaled quantities). Then, at~$s=0$, the~coefficient of restitution is abruptly changed to $\alpha=0.1$ and the evolution toward the corresponding HCS is monitored. We have taken two classes of values of $\alpha_0$: (a) $\alpha_0<1$, corresponding to dissipative inelastic collisions (label I), and~(b) $\alpha_0>1$ \cite{KH04}, corresponding to ``super-elastic'' collisions (label S). More specifically, for~the preparation of the initial state I, we have chosen $\alpha_0=0.29$ and $0.27$ for $d=2$ and $3$, respectively; the state S has been prepared with $\alpha_0=1.29$ and $1.47$ for $d=2$ and $3$, respectively.

Table \ref{table0} displays the values of $a_2$ and $a_3$ corresponding to, in~order of increasing $a_2$, the~initial states $\delta$, M, I, $\Gamma$, and~S.

 \begin{table}[H]%
   \caption{Values of the fourth and sixth cumulants for the initial distributions $\delta$, M, I, $\Gamma$, and~S (see~text).\label{table0}}
\centering
\begin{tabular}{cccccc}
\toprule
&$\bm{\delta}$&\textbf{M}&\textbf{I}&$\bm{\Gamma}$&\textbf{S}\\
\midrule
$a_2(0)$&$\begin{array}{cc}
-0.500&(d=2)\\
-0.400&(d=3)
\end{array}$ &$0$ &$\begin{array}{cc}0.151&(d=2)\\0.111& (d=3)\end{array}$ &$\begin{array}{cc}0.580&(d=2)\\0.580& (d=3)\end{array}$ &$\begin{array}{cc}0.885&(d=2)\\0.792& (d=3)\end{array}$ \\
\midrule
$a_3(0)$&$\begin{array}{cc}
-0.667&(d=2)\\
-0.457&(d=3)
\end{array}$ &$0$ &$\begin{array}{cc}-0.080& (d=2)\\-0.046& (d=3)\end{array}$ &$\begin{array}{cc}-0.062&(d=2)\\-0.149& (d=3)\end{array}$ &$\begin{array}{cc}-4.733&(d=2)\\-2.219& (d=3)\end{array}$ \\
\bottomrule
   \end{tabular}
 \end{table}
\unskip

\section{Formal Expression for $\partial_s\mathcal{D}_{\text{KL}}(\phi\|\phi_\refe)$}
\label{app:A0}
The aim of this appendix is to derive a formal expression for  $\partial_s\mathcal{D}_{\text{KL}}(\phi\|\phi_\refe)$ by following the same steps as in the proof of the conventional $H$-theorem~\cite{GS03}.

Let us consider a generic test function $\psi(\cc)$. By~standard steps, one can easily obtain~\cite{G19}
\bal
\label{A0_1}
\mathcal{J}[\psi]\equiv&\int\dif\cc\,\psi(\cc)I[\cc_1|\phi,\phi]\nn
=&\frac{1}{2}\int \dif\cc_1\int \dif\cc_2\int_+ \dif \s (\cc_{12}\cdot\s)\phi(\cc_1)\phi(\cc_2)
\left[\psi(\cc_1')+\psi(\cc_2')-\psi(\cc_1)-\psi(\cc_2)\right].
\eal
Next, we perform the change of variables $\{\cc_1,\cc_2,\s\}\to \{\cc_1',\cc_2',-\s\}$ and take into account that $\dif\cc_1'\dif\cc_2'=\alpha \dif\cc_1 \dif\cc_2$ and $\cc_{12}'\cdot\s=-\alpha\cc_{12}\cdot\s$ to obtain
\bal
\label{A0_2}
\mathcal{J}[\psi]=&\frac{\alpha^{-2}}{2}\int \dif\cc_1'\int \dif\cc_2'\int_+ \dif \s (\cc_{12}'\cdot\s)\phi(\cc_1)\phi(\cc_2)\left[\psi(\cc_1')+\psi(\cc_2')-\psi(\cc_1)-\psi(\cc_2)\right]\nn
=&\frac{\alpha^{-2}}{2}\int \dif\cc_1\int \dif\cc_2\int_+ \dif \s (\cc_{12}\cdot\s)\phi(\cc_1'')\phi(\cc_2'')\left[\psi(\cc_1)+\psi(\cc_2)-\psi(\cc_1'')-\psi(\cc_2'')\right],
\eal
where in the second equality we have just renamed $\{\cc_1',\cc_2',\cc_1,\cc_2\}\to \{\cc_1,\cc_2,\cc_1'',\cc_2''\}$. Taking the average between Equations \eqref{A0_1} and \eqref{A0_2}, we arrive at
\bal
\label{A0_3}
\mathcal{J}[\psi]=&\frac{1}{4}\int \dif\cc_1\int \dif\cc_2\int_+ \dif \s (\cc_{12}\cdot\s)\Big\{\phi(\cc_1)\phi(\cc_2)\left[\psi(\cc_1')+\psi(\cc_2')-\psi(\cc_1)-\psi(\cc_2)\right]\nn
&-
\frac{\phi(\cc_1'')\phi(\cc_2'')}{\alpha^2}\left[\psi(\cc_1'')+\psi(\cc_2'')-\psi(\cc_1)-\psi(\cc_2)\right]\Big\}.
\eal

Now, we start from the KLD defined by Equation \eqref{KLD} and use the Boltzmann equation \eqref{BE_phi} to get
\begin{equation}
\label{A0_4}
\frac{\kappa}{2}\partial_s\DKL(\phi\|\phi_\refe)=\mathcal{J}\left[\ln\frac{\phi}{\phi_\refe}\right]-\frac{\mu_2}{d}\int \dif\cc\, \ln\frac{\phi(\cc)}{\phi_\refe(\cc)}\frac{\partial}{\partial\cc}\cdot \cc \phi(\cc).
\end{equation}
where we have taken into account that $\phi_\refe(\cc)$  and $\int\dif\cc\,\phi(\cc)=1$ are independent of time. Integration by parts of the second term on the right-hand side of Equation \eqref{A0_4} yields
\begin{equation}
\label{A0_5}
\frac{\kappa}{2}\partial_s\DKL(\phi\|\phi_\refe)=\mathcal{J}\left[\ln\frac{\phi}{\phi_\refe}\right]-{\mu_2}\left[1+\frac{1}{d}\int \dif\cc\, \phi(\cc)\cc\cdot\frac{\partial}{\partial\cc} \ln \phi_\refe(\cc)\right].
\end{equation}
Finally, making use of Equation \eqref{A0_3} with $\psi(\cc)=\ln [\phi(\cc)/\phi_\refe(\cc)]$, we obtain
\bal
\label{A0_6}
\frac{\kappa}{2}\partial_s\DKL(\phi\|\phi_\refe)=&\frac{1}{4}\int \dif\cc_1\int \dif\cc_2\int_+ \dif \s (\cc_{12}\cdot\s)\Bigg[\phi(\cc_1)\phi(\cc_2)\ln\frac{\phi(\cc_1')\phi(\cc_2')\phi_\refe(\cc_1)\phi_\refe(\cc_2)}{\phi(\cc_1)\phi(\cc_2)\phi_\refe(\cc_1')\phi_\refe(\cc_2')}
\nn
&-
\frac{\phi(\cc_1'')\phi(\cc_2'')}{\alpha^2}\ln\frac{\phi(\cc_1'')\phi(\cc_2'')\phi_\refe(\cc_1)\phi_\refe(\cc_2)}{\phi(\cc_1)\phi(\cc_2)\phi_\refe(\cc_1'')\phi_\refe(\cc_2'')}\Bigg]
-\frac{\mu_2}{d}\int \dif\cc\, \phi(\cc)\cc\cdot\frac{\partial}{\partial\cc} \ln \frac{\phi_\refe(\cc)}{\phi_{\text{M}}(\cc)},
\eal
where we have taken into account that $-\int\dif\cc\, \phi(\cc)\cc\cdot\frac{\partial}{\partial\cc} \ln \phi_{\text{M}}(\cc)=2\int\dif\cc\, c^2 \phi(\cc)=d$.

Equation \eqref{A0_6} does not particularly simplify if $\phi_\refe=\phi_\hcs$. However, in~the case $\phi_\refe=\phi_\text{M}$, a somewhat simpler expression can be found. First, the~last term on the right-hand side of Equation \eqref{A0_6} vanishes if $\phi_\refe=\phi_\text{M}$. Second, we can use the decomposition $\mathcal{J}[\ln(\phi/\phi_\text{M})]=\mathcal{J}[\ln \phi]-\mathcal{J}[\ln\phi_\text{M}]$ and take into account that $\ln \phi_\text{M}(\cc)=-c^2+\text{const}$ and, therefore, $\mathcal{J}[\ln\phi_\text{M}]=\mu_2$ [see Equation \eqref{BE_phi}]. As~a consequence,
\bal
\label{A0_7}
\frac{\kappa}{2}\partial_s\DKL(\phi\|\phi_\text{M})=&\frac{1}{4}\int \dif\cc_1\int \dif\cc_2\int_+ \dif \s (\cc_{12}\cdot\s)\Bigg[\phi(\cc_1)\phi(\cc_2)\ln\frac{\phi(\cc_1')\phi(\cc_2')}{\phi(\cc_1)\phi(\cc_2)}
\nn&
-
\frac{\phi(\cc_1'')\phi(\cc_2'')}{\alpha^2}\ln\frac{\phi(\cc_1'')\phi(\cc_2'')}{\phi(\cc_1)\phi(\cc_2)}\Bigg]-\mu_2.
\eal
In the special case of elastic collisions ($\alpha=1$), one has $\mu_2=0$ and $c_{i}''=c_i'$, so that the standard $H$-theorem is recovered, namely
\bal
\label{A0_8}
\left.\frac{\kappa}{2}\partial_s\DKL(\phi\|\phi_\text{M})\right|_{\alpha=1}=&-\frac{1}{4}\int \dif\cc_1\int \dif\cc_2\int_+ \dif \s (\cc_{12}\cdot\s)\left[\phi(\cc_1')\phi(\cc_2')-\phi(\cc_1)\phi(\cc_2)\right]\ln\frac{\phi(\cc_1')\phi(\cc_2')}{\phi(\cc_1)\phi(\cc_2)}\leq 0.
\eal

%%%%%%%%%%%%%%%%%%%%%%%%%%%%%%%%%%%%%%%%%%

%%%%%%%%%%%%%%%%%%%%%%%%%%%%%%%%%%%%%%%%%%
% Citations and References in Supplementary files are permitted provided that they also appear in the reference list here.

%=====================================
% References, variant A: internal bibliography
%=====================================
\reftitle{References}
%\bibliography{biblio}

% The following MDPI journals use author-date citation: Arts, Econometrics, Economies, Genealogy, Humanities, IJFS, JRFM, Laws, Religions, Risks, Social Sciences. For those journals, please follow the formatting guidelines on http://www.mdpi.com/authors/references
% To cite two works by the same author: \citeauthor{ref-journal-1a} (\citeyear{ref-journal-1a}, \citeyear{ref-journal-1b}). This produces: Whittaker (1967, 1975)
% To cite two works by the same author with specific pages: \citeauthor{ref-journal-3a} (\citeyear{ref-journal-3a}, p. 328; \citeyear{ref-journal-3b}, p.475). This produces: Wong (1999, p. 328; 2000, p. 475)

%=====================================
% References, variant B: external bibliography
%=====================================
    %\externalbibliography{yes}
    %\bibliography{D:/Dropbox/Mis_Dropcumentos/bib_files/Granular}

\begin{thebibliography}{999}
\providecommand{\natexlab}[1]{#1}

\bibitem[{Shannon}(1948)]{S48}
{Shannon}, C.E.
\newblock A mathematical theory of communication.
\newblock {\em Bell Syst. Tech. J.} {\bf 1948}, {\em 27},~379--423, doi:10.1002/j.1538-7305.1948.tb01338.x.

\bibitem[Gray(2011)]{G11b}
Gray, R.M.
\newblock {\em Entropy and Information Theory}, 2nd ed.; Springer: New York, NY, USA,
  2011.

\bibitem[Brey and Santos(1992)]{BS92}
Brey, J.J.; Santos, A.
\newblock {Nonequilibrium entropy of a gas}.
\newblock {\em Phys. Rev. A} {\bf 1992}, {\em 45},~8566--8572, doi:10.1103/PhysRevA.45.8566.

\bibitem[Kremer(2014)]{K14}
Kremer, G.M.
\newblock {Thermodynamics and kinetic theory of granular materials}.
\newblock  In \emph{Perspectives and Challenges in Statistical Physics and Complex
  Systems for the Next Decade}; Viswanathan, G.M., Raposo, E.P., {da Luz},
  M.G.E., Eds.; World Scientific: Singapore,  2014; pp. 287--299, doi:10.1142/9789814590143\_0016.

\bibitem[Chapman and Cowling(1970)]{CC70}
Chapman, S.; Cowling, T.G.
\newblock {\em The Mathematical Theory of Non-Uniform Gases}, 3rd ed.; Cambridge
  University Press: Cambridge, UK,  1970.

\bibitem[Garz{\'o} and Santos(2003)]{GS03}
Garz{\'o}, V.; Santos, A.
\newblock {\em Kinetic Theory of Gases in Shear Flows: Nonlinear Transport};
  Fundamental Theories of Physics; Springer: Dordrecht, The Netherlands, 2003.

\bibitem[Kullback and Leibler(1951)]{KL51}
Kullback, S.; Leibler, R.A.
\newblock On Information and Sufficiency.
\newblock {\em Ann. Math. Statist.} {\bf 1951}, {\em 22},~79--86, doi:10.1214/aoms/1177729694.

\bibitem[Kullback(1978)]{K78}
Kullback, S.
\newblock {\em Information Theory and Statistics}; Dover: New York,  NY, USA, 1978.

\bibitem[Santos and Kremer(2012)]{SK12}
Santos, A.; Kremer, G.M.
\newblock Relative Entropy of a Freely Cooling Granular Gas.
\newblock {\em AIP Conf. Proc.} {\bf 2012}, {\em 1501},~1044--1050, doi:10.1063/1.4769657.

\bibitem[{Bettolo Marconi} \em{et~al.}(2013){Bettolo Marconi}, Puglisi, and
  Vulpiani]{BPV13}
{Bettolo Marconi}, U.M.; Puglisi, A.; Vulpiani, A.
\newblock About an {H}-theorem for systems with non-conservative interactions.
\newblock {\em J. Stat. Mech.} {\bf 2013},  {P08003}, doi:10.1088/1742-5468/2013/08/P08003.

\bibitem[{Garc{\'{\i}}a de Soria} \em{et~al.}(2015){Garc{\'{\i}}a de Soria},
  Maynar, Mischler, Mouhot, Rey, and Trizac]{GMMMRT15}
{Garc{\'{\i}}a de Soria}, M.I.; Maynar, P.; Mischler, S.; Mouhot, C.; Rey, T.;
  Trizac, E.
\newblock Towards an {H}-theorem for granular gases.
\newblock {\em J. Stat. Mech.} {\bf 2015},  P11009, doi:10.1088/1742-5468/2015/11/p11009.

\bibitem[Plata and Prados(2017)]{PP17}
Plata, C.A.; Prados, A.
\newblock Global stability and $H$ theorem in lattice models with
  nonconservative interactions.
\newblock {\em Phys. Rev. E} {\bf 2017}, {\em 95},~{052}{121}, doi:10.1103/PhysRevE.95.052121.

\bibitem[Bannerman \em{et~al.}(2011)Bannerman, Sargant, and Lue]{BSL11}
Bannerman, M.N.; Sargant, R.; Lue, L.
\newblock DynamO: A Free $\mathcal{O}(N)$ General Event-Driven Molecular
  Dynamics Simulator.
\newblock {\em J. Comput. Chem.} {\bf 2011}, {\em 32},~3329--3338, doi:10.1002/jcc.21915.

\bibitem[Garz\'o(2019)]{G19}
Garz\'o, V.
\newblock {\em Granular Gaseous Flows. A Kinetic Theory Approach to Granular
  Gaseous Flows}; Springer Nature: Cham, Switzerland,  2019.

\bibitem[Brilliantov and P\"oschel(2004)]{BP04}
Brilliantov, N.V.; P\"oschel, T.
\newblock {\em Kinetic Theory of Granular Gases}; Oxford University Press:
  Oxford, UK,  2004.

\bibitem[Brilliantov and P\"oschel(2000)]{BP00}
Brilliantov, N.; P\"oschel, T.
\newblock Deviation from {Maxwel}l distribution in granular gases with constant
  restitution coefficient.
\newblock {\em Phys. Rev. E} {\bf 2000}, {\em 61},~2809--2812, doi:10.1103/PhysRevE.61.2809.

\bibitem[{van Noije} and Ernst(1998)]{vNE98}
{van Noije}, T.P.C.; Ernst, M.H.
\newblock Velocity distributions in homogeneous granular fluids: The free and
  the heated case.
\newblock {\em Granul. Matter} {\bf 1998}, {\em 1},~57--64, doi:10.1007/s100350050009.

\bibitem[Montanero and Santos(2000)]{MS00}
Montanero, J.M.; Santos, A.
\newblock Computer simulation of uniformly heated granular fluids.
\newblock {\em Granul. Matter} {\bf 2000}, {\em 2},~53--64, doi:10.1007/s100350050035.

\bibitem[Santos and Montanero(2009)]{SM09}
Santos, A.; Montanero, J.M.
\newblock The second and third {S}onine coefficients of a freely cooling
  granular gas revisited.
\newblock {\em Granul. Matter} {\bf 2009}, {\em 11},~157--168, doi:10.1007/s10035-009-0132-8.

\bibitem[Brey \em{et~al.}(1996)Brey, Ruiz-Montero, and Cubero]{BRC96}
Brey, J.J.; Ruiz-Montero, M.J.; Cubero, D.
\newblock Homogeneous cooling state of a low-density granular flow.
\newblock {\em \mbox{Phys. Rev. E}} {\bf 1996}, {\em 54},~3664, doi:10.1103/PhysRevE.54.3664.

\bibitem[Ahmad and Puri(2006)]{AP06}
Ahmad, S.R.; Puri, S.
\newblock Velocity distributions in a freely evolving granular gas.
\newblock {\em Europhys. Lett.} {\bf 2006}, {\em 75},~56--62, doi:10.1209/epl/i2006-10071-3.

\bibitem[Ahmad and Puri(2007)]{AP07}
Ahmad, S.R.; Puri, S.
\newblock Velocity distributions and aging in a cooling granular gas.
\newblock {\em Phys. Rev. E} {\bf 2007}, {\em 75},~031302, doi:10.1103/PhysRevE.75.031302.

\bibitem[Yu \em{et~al.}(2020)Yu, Schr\"oter, and Sperl]{YSS20}
Yu, P.; Schr\"oter, M.; Sperl, M.
\newblock Velocity Distribution of a Homogeneously Cooling Granular Gas.
\newblock {\em \mbox{Phys. Rev. Lett.}} {\bf 2020}, {\em 124},~208007, doi:10.1103/PhysRevLett.124.208007.

\bibitem[Bobylev \em{et~al.}(2003)Bobylev, Cercignani, and Toscani]{BCT03}
Bobylev, A.V.; Cercignani, C.; Toscani, G.
\newblock Proof of an asymptotic property of self-similar solutions of the
  {Boltzmann} equation for granular materials.
\newblock {\em J. Stat. Phys.} {\bf 2003}, {\em 111},~403--417, doi:10.1023/A:1022273528296.

\bibitem[Bisi \em{et~al.}(2006)Bisi, Carrillo, and Toscani]{BCT06}
Bisi, M.; Carrillo, J.A.; Toscani, G.
\newblock Decay Rates in Probability Metrics Towards Homogeneous Cooling States
  for the Inelastic {Maxwell} Model.
\newblock {\em J. Stat. Phys.} {\bf 2006}, {\em 124},~625--653, doi:10.1007/s10955-006-9035-9.

\bibitem[Bolley and Carrillo(2007)]{BC07}
Bolley, F.; Carrillo, J.A.
\newblock Tanaka Theorem for Inelastic {Maxwell} Models.
\newblock {\em Commun. Math. Phys.} {\bf 2007}, {\em 276},~287--314, doi:10.1007/s00220-007-0336-x.

\bibitem[Carrillo and Toscani(2007)]{CT07}
Carrillo, J.A.; Toscani, G.
\newblock Contractive probability metrics and asymptotic behavior of
  dissipative kinetic equations.
\newblock {\em Riv. Mat. Univ. Parma} {\bf 2007}, {\em 6},~75--198.

\bibitem[Carlen \em{et~al.}(2009)Carlen, Carrillo, and Carvalho]{CCC09}
Carlen, E.A.; Carrillo, J.A.; Carvalho, M.C.
\newblock Strong Convergence towards homogeneous cooling states for dissipative
  {Maxwell} models.
\newblock {\em Ann. I. H. Poincar\'e } {\bf 2009}, {\em 26},~167--1700, doi:10.1016/j.anihpc.2008.10.005.

\bibitem[Brilliantov and P\"oschel(2006{\natexlab{a}})]{BP06}
Brilliantov, N.; P\"oschel, T.
\newblock Breakdown of the {Sonine} expansion for the velocity distribution of
  granular gases.
\newblock {\em Europhys. Lett.} {\bf 2006}, {\em 74},~424--430, doi:10.1209/epl/i2005-10555-6;
 {Erratum}: %Authors: Old Ref. 30 combined
 {\bf 2006}, {\em 75},~188--188, doi:10.1209/epl/i2006-10099-3.

\bibitem[Noskowicz \em{et~al.}(2007)Noskowicz, Bar-Lev, Serero, and
  Goldhirsch]{NBSG07}
Noskowicz, S.H.; Bar-Lev, O.; Serero, D.; Goldhirsch, I.
\newblock Computer-aided kinetic theory and granular gases.
\newblock {\em EPL} {\bf 2007}, {\em 79},~{60}{001}, doi:10.1209/0295-5075/79/60001.

\bibitem[Brito and Ernst(1998)]{BE98}
Brito, R.; Ernst, M.H.
\newblock Extension of {H}aff's cooling law in granular flows.
\newblock {\em Europhys. Lett.} {\bf 1998}, {\em 43},~497--502, doi:10.1209/epl/i1998-00388-9.

\bibitem[Goldshtein and Shapiro(1995)]{GS95}
Goldshtein, A.; Shapiro, M.
\newblock Mechanics of collisional motion of granular materials. {P}art 1.
  {G}eneral hydrodynamic equations.
\newblock {\em J. Fluid Mech.} {\bf 1995}, {\em 282},~75--114, doi:10.1017/S0022112095000048.

\bibitem[Coppex \em{et~al.}(2003)Coppex, Droz, Piasecki, and Trizac]{CDPT03}
Coppex, F.; Droz, M.; Piasecki, J.; Trizac, E.
\newblock On the first {S}onine correction for granular gases.
\newblock {\em Physical A} {\bf 2003}, {\em 329},~114--126, doi:10.1016/S0378-4371(03)00593-4.

\bibitem[Abramowitz and Stegun(1972)]{AS72}
Abramowitz, M.; Stegun, I.A.  (Eds.)
\newblock {\em Handbook of Mathematical Functions};  Dover: New York,  NY, USA, 1972.

\bibitem[Bird(1994)]{B94}
Bird, G.A.
\newblock {\em Molecular Gas Dynamics and the Direct Simulation of Gas Flows};
  Clarendon: Oxford, UK,  1994.

\bibitem[Brey \em{et~al.}(1998)Brey, Dufty, Kim, and Santos]{BDKS98}
Brey, J.J.; Dufty, J.W.; Kim, C.S.; Santos, A.
\newblock Hydrodynamics for granular flow at low density.
\newblock {\em Phys. Rev. E} {\bf 1998}, {\em 58},~4638--4653, doi:10.1103/PhysRevE.58.4638.

\bibitem[Boltzmann(1995)]{B95}
Boltzmann, L.
\newblock {\em Lectures on Gas Theory}; Dover: New York,  NY, USA, 1995.

\bibitem[Maynar and Trizac(2011)]{MT11}
Maynar, P.; Trizac, E.
\newblock Entropy of Continuous Mixtures and the Measure Problem.
\newblock {\em Phys. Rev. Lett.} {\bf 2011}, {\em 106},~160603, doi:10.1103/PhysRevLett.106.160603.

\bibitem[Mischler \em{et~al.}(2006)Mischler, Mouhot, and {Rodriguez
  Ricard}]{MMR06}
Mischler, S.; Mouhot, C.; {Rodriguez Ricard}, M.
\newblock Cooling Process for Inelastic {B}oltzmann Equations for Hard Spheres,
  {P}art I: The {C}auchy Problem.
\newblock {\em J. Stat. Phys.} {\bf 2006}, {\em 124},~655--702, doi:10.1007/s10955-006-9096-9.

\bibitem[Mischler and Mouhot(2006)]{MM06}
Mischler, S.; Mouhot, C.
\newblock Cooling Process for Inelastic {B}oltzmann Equations for Hard Spheres,
  {P}art II: Self-Similar Solutions and Tail Behavior.
\newblock {\em J. Stat. Phys,} {\bf 2006}, {\em 124},~703--746, doi:10.1007/s10955-006-9097-8.

\bibitem[Mischler and Mouhot(2009)]{MM09}
Mischler, S.; Mouhot, C.
\newblock Stability, Convergence to Self-Similarity and Elastic Limit for the
  {B}oltzmann Equation for Inelastic Hard Spheres.
\newblock {\em Commun. Math. Phys.} {\bf 2009}, {\em 288},~431--502, doi:10.1007/s00220-009-0773-9.

\bibitem[Pettersson(2004)]{P04}
Pettersson, R.
\newblock On Solutions to the Linear {B}oltzmann Equation for Granular Gases.
\newblock {\em Transp. Theory Stat. Phys.} {\bf 2004}, {\em 33},~527--543, doi:10.1081/TT-200053937.

\bibitem[Esipov and P{\"o}schel(1997)]{EP97}
Esipov, S.E.; P{\"o}schel, T.
\newblock The granular phase diagram.
\newblock {\em J. Stat. Phys.} {\bf 1997}, {\em 86},~1385--1395, doi:10.1007/BF02183630.

\bibitem[Kuninaka and Hayakawa(2004)]{KH04}
Kuninaka, H.; Hayakawa, H.
\newblock Anomalous Behavior of the Coefficient of Normal Restitution in
  Oblique Impact.
\newblock {\em Phys. Rev. Lett.} {\bf 2004}, {\em 93},~{154}{301}, doi:10.1103/PhysRevLett.93.154301.

\end{thebibliography}

%%%%%%%%%%%%%%%%%%%%%%%%%%%%%%%%%%%%%%%%%%
\end{document}